\documentclass[10pt,journal,doublecolumn,twoside]{IEEEtran}
\IEEEoverridecommandlockouts

\usepackage{subfigure}
\usepackage{colortbl}
\usepackage{bm}

\usepackage{graphicx}  
\usepackage{url}       
\usepackage{rotating}
\usepackage{amsmath}   
\usepackage{cite}

\usepackage{amsfonts,amssymb}

\usepackage{stfloats}

\usepackage{cases}
\usepackage{algorithm}
\usepackage{multirow}
\usepackage{algorithmic}

\newcommand{\myvec}[1]%
{\stackrel{\raisebox{-2pt}[0pt][0pt]
{\small$\rightharpoonup$}}{#1}}

\newcommand{\ls}[1]
    {\dimen0=\fontdimen6\the\font
     \lineskip=#1\dimen0
     \advance\lineskip.5\fontdimen5\the\font
     \advance\lineskip-\dimen0
     \lineskiplimit=.9\lineskip
     \baselineskip=\lineskip
     \advance\baselineskip\dimen0
     \normallineskip\lineskip
     \normallineskiplimit\lineskiplimit
     \normalbaselineskip\baselineskip
     \ignorespaces
    }

\begin{document}
\setlength{\columnsep}{0.24in}

\title{Radio Vortex Wireless Communications With Non-Coaxial UCA Transceiver}

\author{Haiyue Jing and Wenchi Cheng

\IEEEauthorblockA{State Key Laboratory of Integrated
Services Networks, Xidian University, Xi'an, China\\
E-mail: \{\emph{hyjing@stu.xidian.edu.cn}, \emph{wccheng@xidian.edu.cn}\} }
\thanks{This work was supported in part by the National Natural Science Foundation of China (No. 61771368) and the Young Elite Scientists Sponsorship Program by CAST (2016QNRC001).}
}


%


\maketitle
\thispagestyle{empty}
\pagestyle{empty}

\begin{abstract}
In the past decade, more and more researchers have concentrated on orbital-angular-momentum (OAM) based radio vortex wireless communications, which is expected to provide orthogonality among different OAM-modes. The uniform circular array (UCA) is considered as one promising antenna structure for OAM based radio vortex wireless communications. However, most studies regarding UCA focus on the scenario where the transmit and receive UCAs are aligned with each other. In this paper, we investigate the radio vortex wireless communications with non-coaxial UCA, i.e., the UCA transceivers are parallel but non-coaxial. We study the channel model and develop the mode-decomposition scheme to decompose the OAM-modes. Then, we discuss the impact of included angles on the channel model under non-coaxial scenario. Numerical results are presented to evaluate our developed scheme and show that the spectrum efficiency of the non-coaxial UCA transceiver in some cases is larger than that of the aligned UCA transceiver based radio vortex wireless communications.
\end{abstract}


\section{Introduction}

\IEEEPARstart{T}{he} plane wave based wireless communications have becoming more and more matured, along with the well utilization of the traditional resources such as time and frequency~\cite{5G_Jeff}. To further increase the spectrum efficiency, an efficient way is to explore other dimensional resources. During the past decade, orbital-angular-momentum (OAM), which is a kind of wavefront with helical phase front, has been attracted much attentions because different OAM-modes and the OAM-modes are orthogonal with each other~\cite{Tamburini_2013_Encoding,Yan_2014_High,Freedom_2015}. Thus, using multiple OAM-modes for information transmission are expected to increase the spectrum efficiency for wireless communications.


So far, many researches and experiments have been conducted to verify the feasibility of OAM based radio vortex wireless communications. The authors of ~\cite{Yan_2014_High} demonstrated a 32 Gbps millimeter-wave link using four OAM-modes on each of two polarizations for data transmission. The authors of ~\cite{dual_channel} experimentally demonstrated a 60 GHz wireless communications link using two OAM-modes. The authors of~\cite{mode_divi_2017} theoretically and experimentally concluded that mode division multiplexing using OAM can reduce the receiver complexity and achieve high spectrum efficiency. In addition, the authors of~\cite{OAM_JSAC_Ge,radio_capa_2015} held that OAM can be used in future wireless broadband communications because of its potential ability to achieve high spectrum efficiency. Apart from the radio vortex wireless communications, OAM based free-space optical communications has also been extensively studied~\cite{optical_OAM_2016,twist_OAM_2016}.

Specifically, uniform circular array (UCA) based OAM is considered as one promising antenna architecture for radio vortex wireless communications because of its flexibility in transmitting multiple OAM beams with different OAM-modes~\cite{OAM_study_2010,phased_OAM_2014}. The schemes of OAM with index modulation were proposed to achieve much better error performance than the OAM based mode division multiplexing schemes~\cite{OAM_IM_2018}. The authors of~\cite{mode_hopping_2018} proposed a mode-hopping scheme within the narrow frequency band for anti-jamming in wireless communications, which can achieve the same anti-jamming results as compared with the conventional wideband frequency-hopping schemes. The authors of~\cite{concentric_2017} proposed concentric UCAs system where the transmit and receive UCAs are aligned with each other to increase the spectrum efficiency.

For the UCA antenna structure, it is now highly demanded for strict alignment between the transmit and receive UCAs in current researches.
However, for wireless communications, it is not practical to maintain the transceiver aligned with each other. If the transmit and receive UCAs are non-coaxial with each other, the phase of received signal contains not only the phase of OAM-mode, but also the phase turbulence due to unequal distance transmission at different places of the receiver~\cite{OAM_magazine_2018}, which challenges the efficient receiving for radio vortex wireless communications. Thus, a question is raised that how to decompose the OAM beams with multiple OAM-modes when the transmit and receive UCAs are non-coaxial.

To overcome the above-mentioned problem, in this paper we derive the mathematical model to characterize channel of non-coaxial radio vortex wireless communications. Based on the channel model, we develop the mode-decomposition scheme to obtain the signal corresponding to each OAM-mode. Also, the impact of concluded angles on the channel model is discussed. We conduct extensive simulations to validate and evaluate that the spectrum efficiency of the non-coaxial UCA transceiver under some non-coaxial cases is larger than that of the aligned UCA transceiver in radio vortex wireless communications.


The rest of this paper is organized as follows. Section~\ref{sec:sys} gives the non-coaxial UCA transceiver based radio vortex wireless communications model. Section~\ref{sec:sche} investigates the channel model and develops the mode-decomposition scheme to obtain the receive signal corresponding to each OAM-mode. Section~\ref{sec:nume} evaluates our developed scheme and discusses the channel amplitude gains versus the included angles under non-coaxial scenario. The paper concludes with Section~\ref{sec:conc}.

\section{The Non-Coaxial UCAs Based System Model for Radio Vortex Wireless Communication}\label{sec:sys}

\begin{figure}
\centering
\includegraphics[width=0.40\textwidth, angle=90]{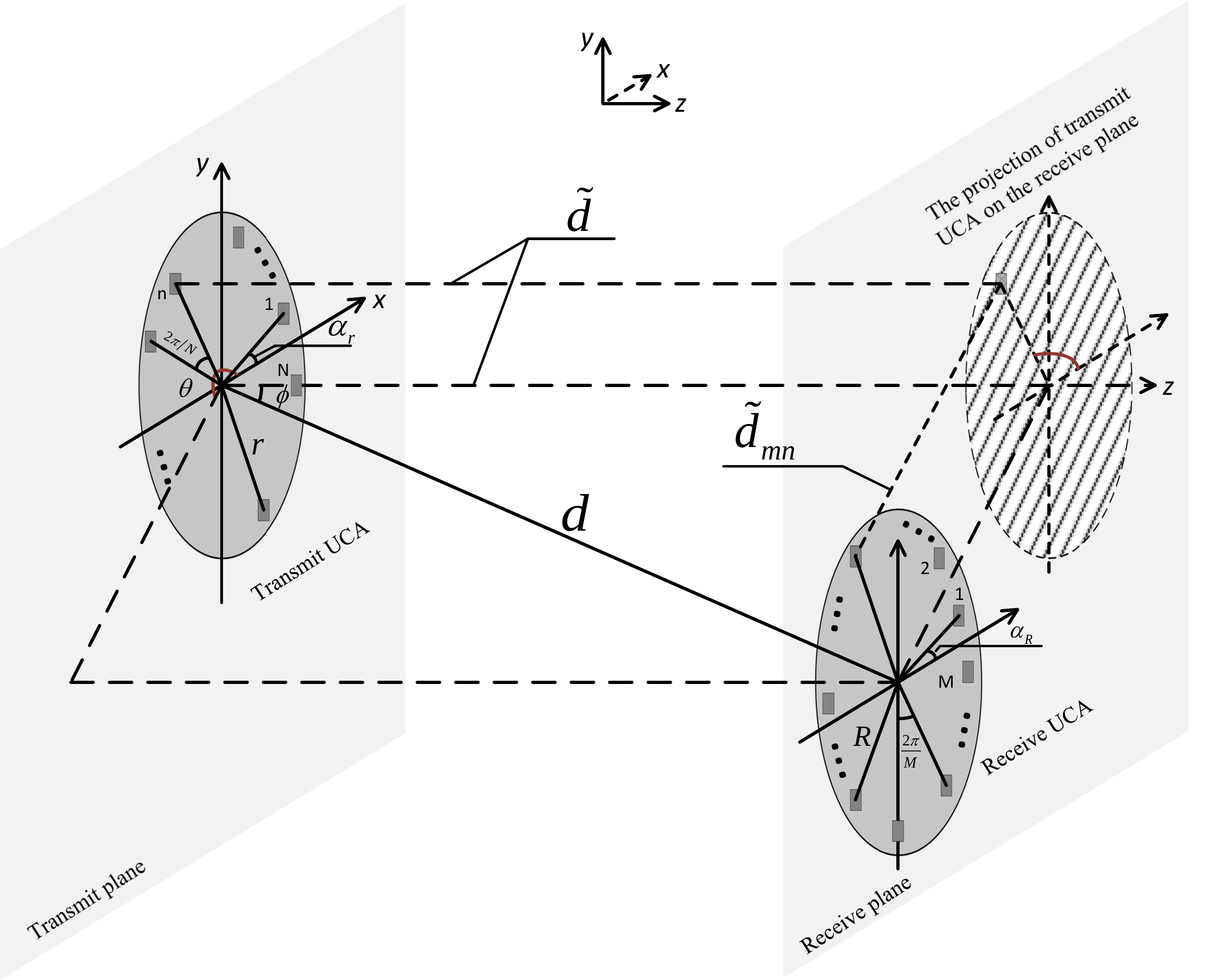}
\caption{The system model for the non-coaxial UCA transceiver based radio vortex wireless communications.}
\label{fig:non_align}
\end{figure}


Figure~\ref{fig:non_align} depicts the system model for the non-coaxial UCA transceiver based radio vortex wireless communications, where the transmit and receive UCAs are non-coaxial and the size of them can be different. The planes corresponding to the transmit and the receive UCAs are the transmit plane and the receive plane, respectively. The projection of the transmit UCA is on the receive plane. We denoted by $d$ the distance from the center of the transmit UCA to the center of the receive UCA. The transmit and receive UCAs are equipped with $N$ array-elements and $M$ array-elements, respectively. For the transmit UCA, the array-elements, which are fed with the same input signal but with a successive delay from array-element to array-element such that after a full turn the phase has been incremented by an integer multiple $l$ of $2\pi$, are uniformly around the perimeter of the circle and $l$ represents the number of topological charges, i.e., the number of OAM-modes. For the receive UCA, the array-elements are also uniformly around perimeter of the circle. We denote by $\alpha_{r}$ and $\alpha_{R}$ the angles between the phase angle of the first array-element and zero radian corresponding to the transmit and receive UCAs, respectively. The parameter $\theta$ denotes the included angle
between $x$-axis and the projection of the line from the center of the transmit UCA to the center of the receive UCA on the transmit plane. Also, $\phi$ denotes the included angle between $z$-axis and the line from the center of the transmit UCA to the center of the receive UCA. The notation $\widetilde{d}$ represents the distance between the center of the transmit UCA and the center of the receive UCA. The parameter $\widetilde{d}_{mn}$ is the distance between the projection of the $n$th ($1\leq n\leq N$) transmit array-element on the receive plane and the $m$th ($1\leq m\leq M$) receive array-element. We also denote by $r$ the radius of transmit UCA and $R$ the radius of receive UCA. In the following, we derive the channel model and develop the mode-decomposition scheme to obtain the signal corresponding to each OAM-mode. 

\section{Channel Model and the Mode-Decomposition Scheme for Non-Coaxial UCAs Based Radio Vortex Wireless Communications}\label{sec:sche}

The signal at the $n$th array-element on the transmit UCA, denoted by $x_{n}$, is given as follows:
\begin{eqnarray}
x_{n} &=& \sum_{l=\left\lfloor\frac{2-N}{2}\right\rfloor}^{\left\lfloor N/2\right\rfloor} \frac{1}{\sqrt{N}} s_{l}e^{j(\varphi_{n}+\alpha_{r}) l}\nonumber \\
&=&\sum_{l=\left\lfloor\frac{2-N}{2}\right\rfloor}^{\left\lfloor N/2\right\rfloor} \frac{1}{\sqrt{N}} s_{l}e^{j\left[\frac{2\pi(n-1)}{N}+\alpha_{r}\right]l},
\end{eqnarray}
where $(\varphi_{n}+\alpha_{r})$ is the azimuthal angle, defined as the angular position on a plane perpendicular to the axis of propagation, corresponding to the $n$th array-element on the transmit UCA. $\varphi_{n}=2\pi(n-1)/N$ is the basic angle for the transmit UCA. The symbol $s_{l}$ denotes the signal on the $l$th OAM-mode of the transmit UCA. $l$ $\left[\left\lfloor (2-N)/2\right\rfloor \leq l \leq \left\lfloor N/2\right\rfloor\right]$ is the OAM-mode number.
When $\kappa<0$, $\left\lfloor \kappa \right\rfloor$ represents the smallest integer which is greater than or equal to $\kappa$. When $\kappa \geq 0$, $\left\lfloor \kappa \right\rfloor$ represents the largest integer which is less than or equal to $\kappa$. 

We denote by $h_{mn}$ the channel gain from the $n$th array-element on the transmit UCA to the $m$th array-element on the receive UCA. Then, $h_{mn}$ can be written as follows~\cite{Is_OAM_2012}:
\begin{eqnarray}\label{channel_1}
h_{mn}=\frac{\beta \lambda e^{-j\frac{2\pi}{\lambda}d_{mn}}}{4\pi d_{mn}},
\end{eqnarray}
where $\beta$ denotes the combination of all the relevant constants such as attenuation and phase rotation caused by antennas and their patterns on both sides. The parameter $d_{mn}$ is the distance between the $n$th array-element on the transmit UCA and the $m$th array-element on the receive UCA.
We denote by $\widetilde{d}_{mn}$ the distance between the projection of the $n$th array-element on the transmit UCA in the receive plane and the $m$th array-element on the receive UCA. Since the transmit UCA and the projection of transmit UCA are aligned well with each other, the distance $\widetilde{d}_{mn}$ should be derived first if we want to obtain $d_{mn}$.

In terms of rectangular coordinates, the coordinate of the $n$th array-element on the transmit UCA is $(r\cos(\varphi_{n}+\alpha_{r}), r\sin(\varphi_{n}+\alpha_{r}), 0)$. The coordinate corresponding to the center of the receive UCA is $(d\sin \phi \cos \theta, d\sin \phi \sin \theta, d\cos \phi)$. Then, we have $\widetilde{d}=d\cos \phi$. Thus, the coordinate related to the projection of the $n$th array-element on the transmit UCA in the receive plane is $(r\cos(\varphi_{n}+\alpha_{r}), r\sin(\varphi_{n}+\alpha_{r}), d\cos \phi)$ and the coordinate of the $m$th array-element on the receive UCA is $(R\cos(\psi_{m}+a_{R})-d\sin \phi \cos \theta, R\sin(\psi_{m}+a_{R})-d\sin \phi \sin \theta, d\cos \phi)$. Similar to the azimuthal angle at the transmit UCA, $\psi_{m}+a_{R}$ is the azimuthal angle of the $m$th array-element on the receive UCA and $\psi_{m}=2\pi(m-1)/M$ is the basic angle for the receive UCA. Therefore, $\widetilde{d}_{mn}$ is derived in Eq.~\eqref{eq:tilde_d_mn}.
\begin{figure*}
\begin{eqnarray}\label{eq:tilde_d_mn}
&&\hspace{-1.2cm}\widetilde{d}_{mn}=\sqrt{\left[R\cos(\psi_{m}+a_{R})-d\sin \phi \cos \theta - r\cos(\varphi_{n}+\alpha_{r})\right]^{2}+\left[R\sin(\psi_{m}+a_{R})-d\sin \phi \sin \theta - r\sin(\varphi_{n}+\alpha_{r})\right]^{2}}\nonumber \\[0.3cm]
&&\hspace{-1.2cm}= \sqrt{R^{2}+r^{2}+d^{2}\sin^{2}\phi-2rR\cos(\psi_{m}+a_{R}-\varphi_{n}-\alpha_{r})-2Rd \sin \phi\cos(\psi_{m}+a_{R}- \theta)+2rd\sin \phi\cos(\varphi_{n}+\alpha_{r}-\theta)}.
\end{eqnarray}
\hrulefill
\end{figure*}
\begin{figure*}
\begin{eqnarray}\label{eq:d_mn}
\!\!\!\!\!\!\!\!\!\!\!\!d_{mn}\!\!\!\!&=&\!\!\!\!\sqrt{\widetilde{d}_{mn}^{2}+\widetilde{d}^2}\nonumber \\
\!\!\!\!&=&\!\!\!\! \sqrt{R^{2}+r^{2}+d^{2}-2rR\cos(\psi_{m}+a_{R}-\varphi_{n}-\alpha_{r})-2Rd \sin \phi\cos(\psi_{m}+a_{R}- \theta)+2rd\sin \phi\cos(\varphi_{n}+\alpha_{r}-\theta)}\nonumber \\[0.3cm]
\!\!\!\!&=&\!\!\!\!\sqrt{R^{2}+r^{2}+d^{2}}\sqrt{1-\frac{2rR\cos(\psi_{m}+a_{R}-\varphi_{n}-\alpha_{r})+2Rd \sin \phi\cos(\psi_{m}+a_{R}- \theta)-2rd\sin \phi\cos(\varphi_{n}+\alpha_{r}-\theta)}{R^{2}+r^{2}+d^{2}}} \nonumber \\[0.3cm]
\!\!\!\!&\approx &\!\!\!\!\sqrt{R^{2}+r^{2}+d^{2}}-\frac{rR\cos(\psi_{m}+a_{R}-\varphi_{n}-\alpha_{r})+Rd \sin \phi\cos(\psi_{m}+a_{R}- \theta)-rd\sin \phi\cos(\varphi_{n}+\alpha_{r}-\theta)}{\sqrt{R^{2}+r^{2}+d^{2}}}.
\end{eqnarray}
\hrulefill
\end{figure*}
\setcounter{equation}{6}
\begin{figure*}
\begin{eqnarray}\label{eq:h_mn}
&&\hspace{-1.2cm}h_{mn}=\underbrace{\frac{\beta \lambda}{4\pi \sqrt{d^2+r^2+R^2}}\exp\left(\frac{-j2\pi\sqrt{d^2+r^2+R^2}}{\lambda}\right) \exp\left[\frac{j2\pi Rd \sin \phi\cos(\psi_{m}+a_{R}- \theta)}{\lambda \sqrt{d^2+r^2+R^2}} \right]}_{A_m} \nonumber \\[0.3cm]
&&\hspace{2cm}\times \exp\left[-j\underbrace{\frac{2\pi r \sqrt{R^2+d^2 \sin^{2}\phi-2Rd\sin\phi\cos(\psi_{m}+a_{R}- \theta)}  }{\lambda \sqrt{d^2+r^2+R^2}}}_{B_m} \sin(\varphi_{n}+\alpha_{r}-\psi_{m}-a_{R}+\zeta_m)    \right].
\end{eqnarray}
\hrulefill
\end{figure*}

Then, using the Pythagorean theorem we can derived $d_{mn}$ in Eq.~\eqref{eq:d_mn}.
Because $d\gg R$ and $d\gg r$, we have $d_{mn}\approx \sqrt{d^2+R^2+r^2}$ for the denominator in Eq.~\eqref{channel_1}. For $d_{mn}$ in the numerator, which is part of the item $\exp\left(-j2\pi d_{mn}/\lambda\right)$, we can approximate it using $\sqrt{1-2x}\approx 1-x$ when $x$ is very close to zero as shown in Eq.~\eqref{eq:d_mn}.

In Eq.~\eqref{eq:d_mn}, the item \big[$2rR\cos(\psi_{m}+a_{R}-\varphi_{n}-\alpha_{r})-2rd\sin \phi\cos(\varphi_{n}+\alpha_{r}-\theta)$\big] can be derived as follows:
\setcounter{equation}{4}
\begin{eqnarray}\label{eq:approxiamte}
&&\hspace{-0.7cm} rR\cos(\psi_{m}+a_{R}-\varphi_{n}-\alpha_{r})-rd\sin \phi\cos(\varphi_{n}+\alpha_{r}-\theta)\nonumber \\
&&\hspace{-0.7cm}=\left[Rr\!-\!rd\sin\phi\cos(\psi_{m}\!+\!a_{R}\!-\! \theta)\right]\cos(\varphi_{n}\!+\!\alpha_{r}\!-\!\psi_{m}\!-\!a_{R})\nonumber \\ [0.15cm]
&&\hspace{-0.0cm}+rd\sin\phi\sin(\psi_{m}+a_{R}- \theta)\sin(\varphi_{n}+\alpha_{r}-\psi_{m}-a_{R})\nonumber \\[0.15cm]
&&\hspace{-0.7cm}= r\sqrt{R^2+d^2 \sin^{2}\phi-2Rd\sin\phi\cos(\psi_{m}+a_{R}- \theta)}\nonumber \\
&&\hspace{2.2cm}\times \sin(\varphi_{n}+\alpha_{r}-\psi_{m}-a_{R}+\zeta_{m}),
\end{eqnarray}
\pagebreak
where
\begin{eqnarray}\left\{
\begin{array}{lll}
\sin\zeta_{m}=\frac{R-d\sin\phi\cos(\psi_{m}+a_{R}- \theta)}{\sqrt{R^2+d^2 \sin^{2}\phi-2Rd\sin\phi\cos(\psi_{m}+a_{R}- \theta)}};\\[0.3cm]
\cos\zeta_m=\frac{d\sin\phi\sin(\psi_{m}+a_{R}- \theta)}{\sqrt{R^2+d^2 \sin^{2}\phi-2Rd\sin\phi\cos(\psi_{m}+a_{R}- \theta)}}.
\end{array}\right.
\label{eq:zeta_m}
\end{eqnarray}

Associating Eqs.~\eqref{channel_1},~\eqref{eq:d_mn},~\eqref{eq:approxiamte}, and~\eqref{eq:zeta_m}, the channel gain $h_{mn}$ can be derived as Eq.~\eqref{eq:h_mn}. For convenient expression, we replace some polynomials in Eq.~\eqref{eq:h_mn} with $A_{m}$ and $B_{m}$, respectively, as shown in Eq.~\eqref{eq:h_mn}.


The received signal at the $m$th array-element, denoted by $y_{m}$, can be derived as follows:
\setcounter{equation}{7}
\begin{eqnarray}
y_{m}&=&\sum_{n=1}^{N}\sum_{l=\left\lfloor\frac{2-N}{2}\right\rfloor}^{\left\lfloor N/2\right\rfloor} h_{mn} \frac{1}{\sqrt{N}} s_{l} e^{j\frac{2\pi(n-1)}{N}l}+z_{m}\nonumber \\
&=&\sum_{l=\left\lfloor\frac{2-N}{2}\right\rfloor}^{\left\lfloor N/2\right\rfloor} \widetilde{h}_{ml}s_{l} + z_{m},
\end{eqnarray}
where $z_{m}$ denotes the received noise at the $m$th array-element on the receive UCA and $z_{m}$ is complex Gaussian variable with zero mean and variance $\sigma_{m}^2$. We denote by $\widetilde{h}_{ml}$ the channel gain from the transmit UCA to the $m$th array-element on the receive UCA corresponding to the $l$th OAM-mode. The channel gain $\widetilde{h}_{ml}$ is given by Eq.~\eqref{eq:h_ml}. In Eq.~\eqref{eq:h_ml}, $\varphi$ is the continuous variable related to $\varphi_{n}$ ranging from 0 and $2\pi$. When $n$ is equal to 1,  $\varphi_{n}$ is equal to zero. When $n=N$, $\varphi_{n}$ is equal to $2\pi-2\pi/N$ and very close to $2\pi$. To express conveniently, we denote by $\psi=\alpha_{r}-\psi_{m}-a_{R}+\zeta_m$, which is used in Eq.~\eqref{eq:h_ml}.

\setcounter{equation}{8}
\begin{figure*}
\begin{eqnarray}\label{eq:h_ml}
&&\hspace{-0.8cm}\widetilde{h}_{ml}=\sum\limits_{n=1}^{N} \frac{1}{\sqrt{N}} h_{mn} e^{j\frac{2\pi(n-1)}{N}l}=\sum\limits_{n=1}^{N} \frac{A_{m}}{\sqrt{N}} \exp\left[-B_{m}\sin(\varphi_{n}+\alpha_{r}-\psi_{m}-a_{R}+\zeta_m)  \right] \exp\left[j (\varphi_{n}+\alpha_{r})l  \right]\nonumber \\
&&\hspace{-1.0cm}= \frac{A_{m}}{\sqrt{N}} \exp\left[j(\psi_{m}+a_{R}-\zeta_m)l\right] \sum\limits_{n=1}^{N} \exp\left[j(\varphi_{n}+\alpha_{r}-\psi_{m}-a_{R}+\zeta_m)l\right]\exp\left[ -B_{m}\sin(\varphi_{n}+\alpha_{r}-\psi_{m}-a_{R}+\zeta_m) \right] \nonumber \\
&&\hspace{-1.0cm}\approx  \sqrt{N} A_{m}\exp\left[j(\psi_{m}+a_{R}-\zeta_m)l\right] \frac{1}{2\pi} \int_{0}^{2\pi} \exp[j(\varphi+\underbrace{\alpha_{r}-\psi_{m}-a_{R}+\zeta_m}_{\psi})l]\exp\left[ -B_{m}\sin(\varphi+\alpha_{r}-\psi_{m}-a_{R}+\zeta_m) \right] d \varphi  \nonumber \\
&&\hspace{-1.0cm}=  \sqrt{N} A_{m}\exp\left[j(\psi_{m}+a_{R}-\zeta_m)l\right] \frac{1}{2\pi} \int_{\psi}^{2\pi+\psi} \exp\left[j(\varphi+\psi)l  \right] \exp\left[-B_{m} \sin(\varphi+\psi)\right]d(\varphi+\psi) \nonumber \\
&&\hspace{-1.0cm}=  \sqrt{N} A_{m}\exp\left[j(\psi_{m}+a_{R}-\zeta_m)l\right] \frac{1}{2\pi} \int_{0}^{2\pi} \exp\left[j(\varphi+\psi)l  \right] \exp\left[-B_{m} \sin(\varphi+\psi)\right]d(\varphi+\psi) \nonumber \\[0.2cm]
&&\hspace{-1.0cm}=  \sqrt{N} A_{m}\exp\left[j(\psi_{m}+a_{R}-\zeta_m)l\right]J_{l}\left(B_{m}\right) \nonumber \\[0.2cm]
&&\hspace{-1.0cm}= \underbrace{\frac{\sqrt{N} \beta \lambda}{4\pi \sqrt{d^2+r^2+R^2}}\exp\left(\frac{-j2\pi\sqrt{d^2+r^2+R^2}}{\lambda}\right)}_{h} \exp\left[j(\psi_{m}+a_{R}-\zeta_m)l\right] \nonumber \\
&&\hspace{0.0cm}\times \underbrace{\exp\left[\frac{j2\pi Rd \sin \phi\cos(\psi_{m}+a_{R}- \theta)}{\lambda \sqrt{d^2+r^2+R^2}} \right] J_{l}\left( \frac{2\pi r \sqrt{R^2+d^2 \sin^{2}\phi-2Rd\sin\phi\cos(\psi_{m}+a_{R}- \theta)}  }{\lambda \sqrt{d^2+r^2+R^2}}\right)}_{C_{m,l}}.
\end{eqnarray}
\hrulefill
\end{figure*}

In Eq.~\eqref{eq:h_ml}, the $l$-order Bessel function is given as follows:
\begin{eqnarray}\label{eq:bessel}
J_{l} (\alpha) =\frac{1}{2\pi}\int_{0}^{2\pi} e^{jl\tau} e^{-j\alpha\sin\tau}d\tau.
\end{eqnarray}
Observing Eq.~\eqref{eq:bessel}, we can find that $e^{jl\tau} e^{-j\alpha\sin\tau}$ is the function, of which $\tau$ is the independent variable, with period $2\pi$ when $l$ is an integer. This is the reason why the line 4 of Eq.~\eqref{eq:h_mn} is equal to the line 5 of Eq.~\eqref{eq:h_mn}. Also, we replace some polynomials in Eq.~\eqref{eq:h_mn} with $h$ and $C_{m,l}$, respectively, as shown in Eq.~\eqref{eq:h_mn} for convenient expression. In the following, we show two cases corresponding to different included angles $\theta$ and $\phi$ as follows:

\noindent {\bf Case A:} When $\phi$ is equal to zero, the transmit and receive UCAs are are coaxial and parallel to each other. In this case,
the channel gain $\widetilde{h}_{ml}$ can be rewritten as follows:
\begin{eqnarray}
\widetilde{h}_{ml}=h J_{l}\left(\frac{2\pi rR}{\lambda \sqrt{d^2+r^2+R^2}}\right) e^{j(\psi_{m}+a_{R}-\frac{\pi}{2})l}.
\end{eqnarray}
We can find that the channel amplitude gain $|\widetilde{h}_{ml}|$ only depends on the order of OAM-mode and $\psi_{m}$ is independent on $|\widetilde{h}_{ml}|$ when $r$, $R$, $\lambda$, and $N$ are fixed. Thus, we can consider that the signals carried by different OAM-modes correspond to different channel amplitude gains.

\noindent {\bf Case B:} When $\phi$ is equal to $\pi/2$, the transmit and receive UCAs are both on the transmit plane. In this case, we denote by $\theta=\alpha_{R}$ and we have
\begin{eqnarray}\left\{
\begin{array}{lll}
B_{m}=\frac{2\pi r \sqrt{R^2+d^2 -2Rd\cos\psi_{m}}  }{\lambda \sqrt{d^2+r^2+R^2}}; \\[0.2cm]
C_{m,l}=\exp\left(\frac{j2\pi Rd \cos\psi_{m}}{\lambda \sqrt{d^2+r^2+R^2}} \right)J_{l}\left(B_{m}\right).
\end{array}\right.
\end{eqnarray}
Thus, the channel amplitude gain $|\widetilde{h}_{ml}|$ depends on $\psi_{m}$ apart from the order of OAM-mode.

The error, denoted by $e_{l}$, corresponding to the $l$th OAM-mode for the approximation in Eq.~\eqref{eq:h_ml} is derived as follows:
\begin{eqnarray}
e_{l}=\log_{10}\left[ h C_{m,l} e^{j\left(\psi_{m}+a_{R}-\zeta_m\right)l} - \sum\limits_{n=1}^{N} \frac{1}{\sqrt{N}} h_{mn} e^{j\varphi_{n}l}\right].\nonumber \\
\end{eqnarray}

Then, based on Eq.~\eqref{eq:h_ml}, we can rewrite $y_{m}$ as follows:
\begin{eqnarray}
y_{m}&=&\!\!\!\sum_{l=\left\lfloor\frac{2-N}{2}\right\rfloor}^{\left\lfloor N/2\right\rfloor} \widetilde{h}_{ml} s_{l} \nonumber \\
&=&\!\!\! \sum_{l=\left\lfloor\frac{2-N}{2}\right\rfloor}^{\left\lfloor N/2\right\rfloor} h C_{m,l} s_{l} e^{j\left[\frac{2\pi (m-1)}{M}+a_{R}-\zeta_m\right]l}+z_{m},
\end{eqnarray}

To recover the transmit signal corresponding to the $l_{0}$th ($\left\lfloor(2-N)/2\right\rfloor \leq l_{0} \leq \left\lfloor N/2\right\rfloor$) OAM-mode, the received signal $y_{m}$ is multiplied with the item $C_{m,l_{0}}^{-1}\exp[-j(\psi_{m}+a_{R}-\zeta_m)l_{0}]$. Then, we denote by $y_{m, l_{0}}$ the received signal related to $l_{0}$ at the $m$th array-element on the receive UCA and $y_{m, l_{0}}$ can be written as follows:
\begin{eqnarray}
y_{m, l_{0}}&=&y_{m}C_{m,l_{0}}^{-1}e^{-j\left[\frac{2\pi (m-1)}{M}+a_{R}-\zeta_m\right]l}\nonumber \\
&=& \sum_{l=\left\lfloor\frac{2-N}{2}\right\rfloor}^{\left\lfloor N/2\right\rfloor} h s_{l} e^{j(a_{R}-\zeta_m)(l-l_{0})} e^{j\frac{2\pi (m-1)}{M}(l-l_{0})}\nonumber \\
&&\hspace{1cm}+z_{m}C_{m,l_{0}}^{-1}e^{-j(\psi_{m}+a_{R}-\zeta_m)l_{0}}.
\end{eqnarray}
Therefore, the received signal, denoted by $y_{l_{0}}$, corresponding to the $l_{0}$th OAM-mode of the receive UCA can be derived as follows:
\begin{eqnarray}
y_{l_{0}} \!\!\!\!\!&=&\!\!\!\!\sum_{m=1}^{M} y_{m, l_{0}}\nonumber \\
\!\!\!\!&=&\!\!\!\! \sum_{l=\left\lfloor\frac{2-N}{2}\right\rfloor, l\neq l_{0}}^{\left\lfloor N/2\right\rfloor} h s_{l} e^{j(a_{R}-\zeta_m)(l-l_{0})}\sum_{m=1}^{M} e^{j\frac{2\pi (m-1)}{M}(l-l_{0})}\nonumber \\
\!\!\!\!&&+ \sum_{m=1}^{M}h s_{l_{0}} + \sum_{m=1}^{M} z_{m}C_{m,l_{0}}^{-1}e^{-j(\psi_{m}+a_{R}-\zeta_m)l_{0}} \nonumber \\
\!\!\!\!&=&\!\!\!\!M h s_{l_{0}}+\sum_{m=1}^{M} z_{m}C_{m,l_{0}}^{-1}e^{-j(\psi_{m}+a_{R}-\zeta_m)l_{0}}.
\end{eqnarray}
Traversing the OAM-mode $l_{0}$, we can obtain all the estimated transmit signals. Because $z_{m}$ is the complex Gaussian variable with zero mean and variance $\sigma_{m}^2$, we can derive that $\sum_{m=1}^{M} z_{m}C_{m,l_{0}}^{-1}e^{-j(\psi_{m}+a_{R}-\zeta_m)l_{0}}$ is a complex Gaussian variable with zero mean and variance $\sum_{m=1}^{M} C_{m,l_{0}}^{-2} \sigma_{m}^2$. Then, the spectrum efficiency, denoted by $C_{OAM}$, for the radio vortex wireless communications under the non-coaxial UCA transceiver scenario can be derived as follows:
\begin{eqnarray}
C_{OAM}= \sum_{l=\left\lfloor\frac{2-N}{2}\right\rfloor}^{\left\lfloor N/2\right\rfloor} \log_{2}\left( 1+\frac{M^2 h^2 |s_{l}|^2}{\sum_{m=1}^{M} C_{m,l_{0}}^{-2} \sigma_{m}^2}\right).
\end{eqnarray}

\vspace{5pt}
\section{Performance Evaluations}\label{sec:nume}
\vspace{3pt}
In this section, we evaluate the performance of non-coaxial UCA transceiver based radio vortex wireless communications. First, we evaluate the approximation error corresponding to different OAM-modes. Then, we evaluate the channel amplitude gains for different OAM-modes versus the included angle $\theta$ and $\phi$, respectively.

\begin{figure}
    \centering
    \vspace{0pt}

\centering
\begin{minipage}[t]{0.32\linewidth}
\centering
\centerline{\includegraphics[width=1\linewidth]{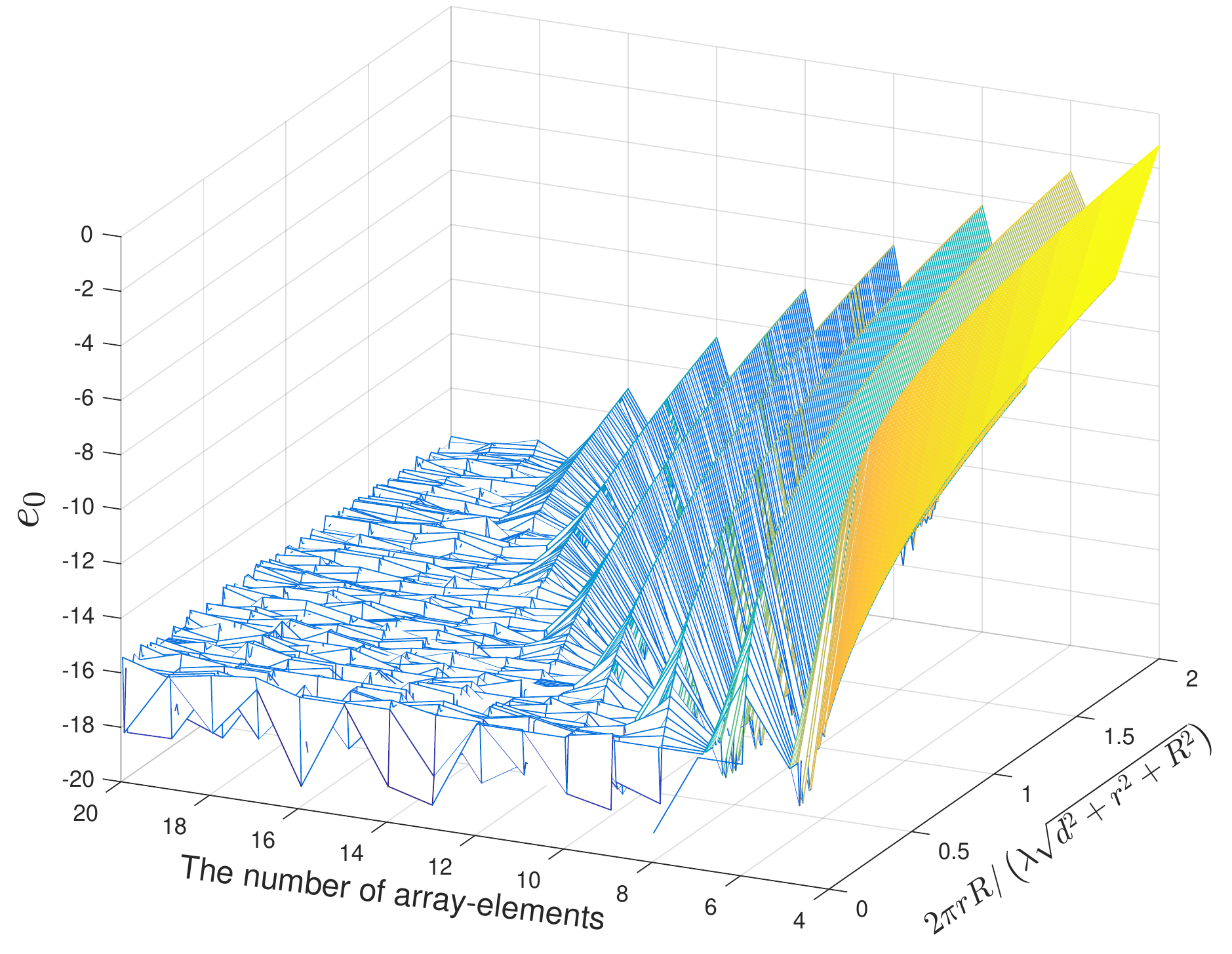}}
\centerline{$l = 0$}
\end{minipage}
\centering
\begin{minipage}[t]{0.32\linewidth}
\centering
\centerline{\includegraphics[width=1\linewidth]{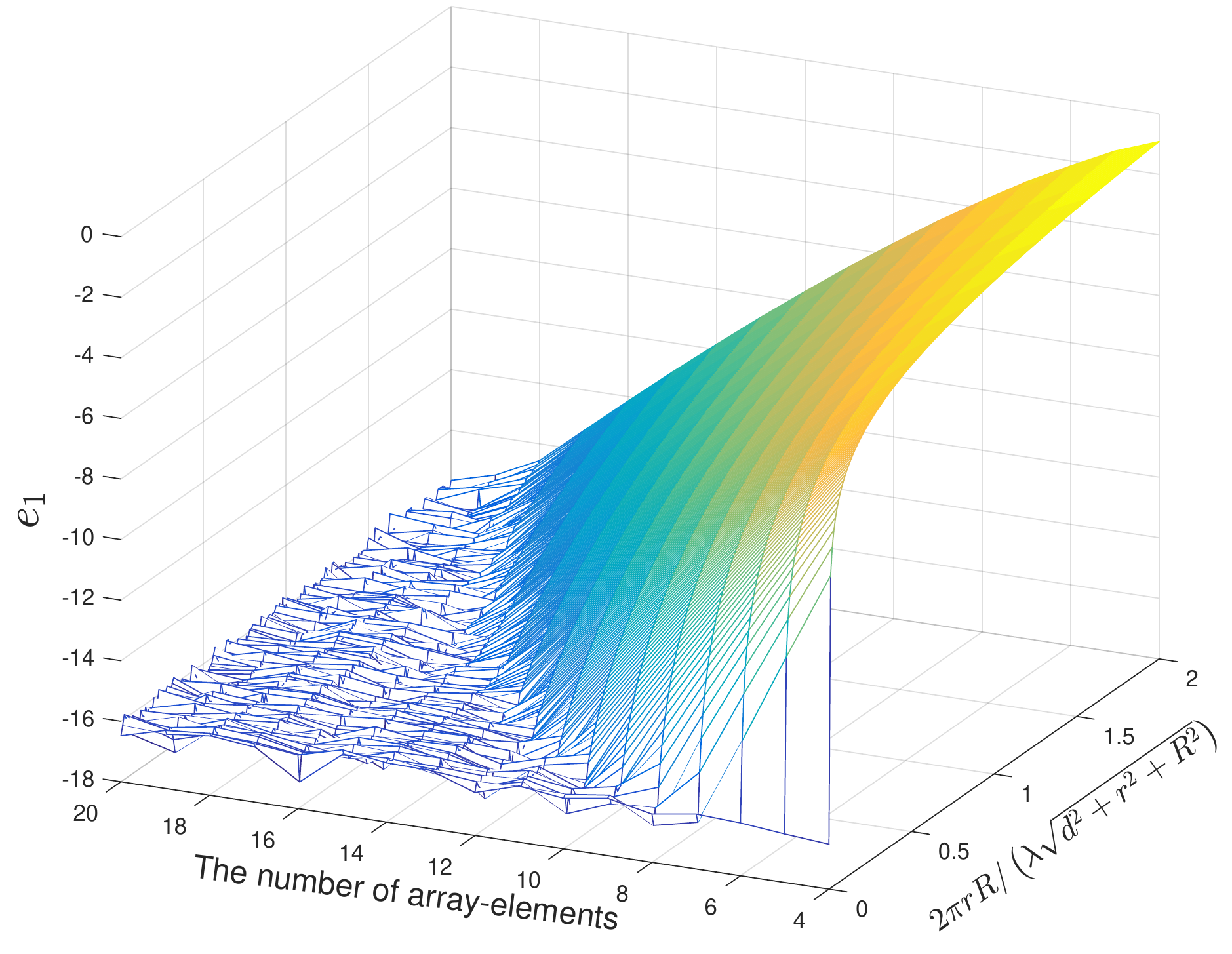}}
\centerline{$l = 1$}
\end{minipage}
\centering
\begin{minipage}[t]{0.32\linewidth}
\centering
\centerline{\includegraphics[width=1\linewidth]{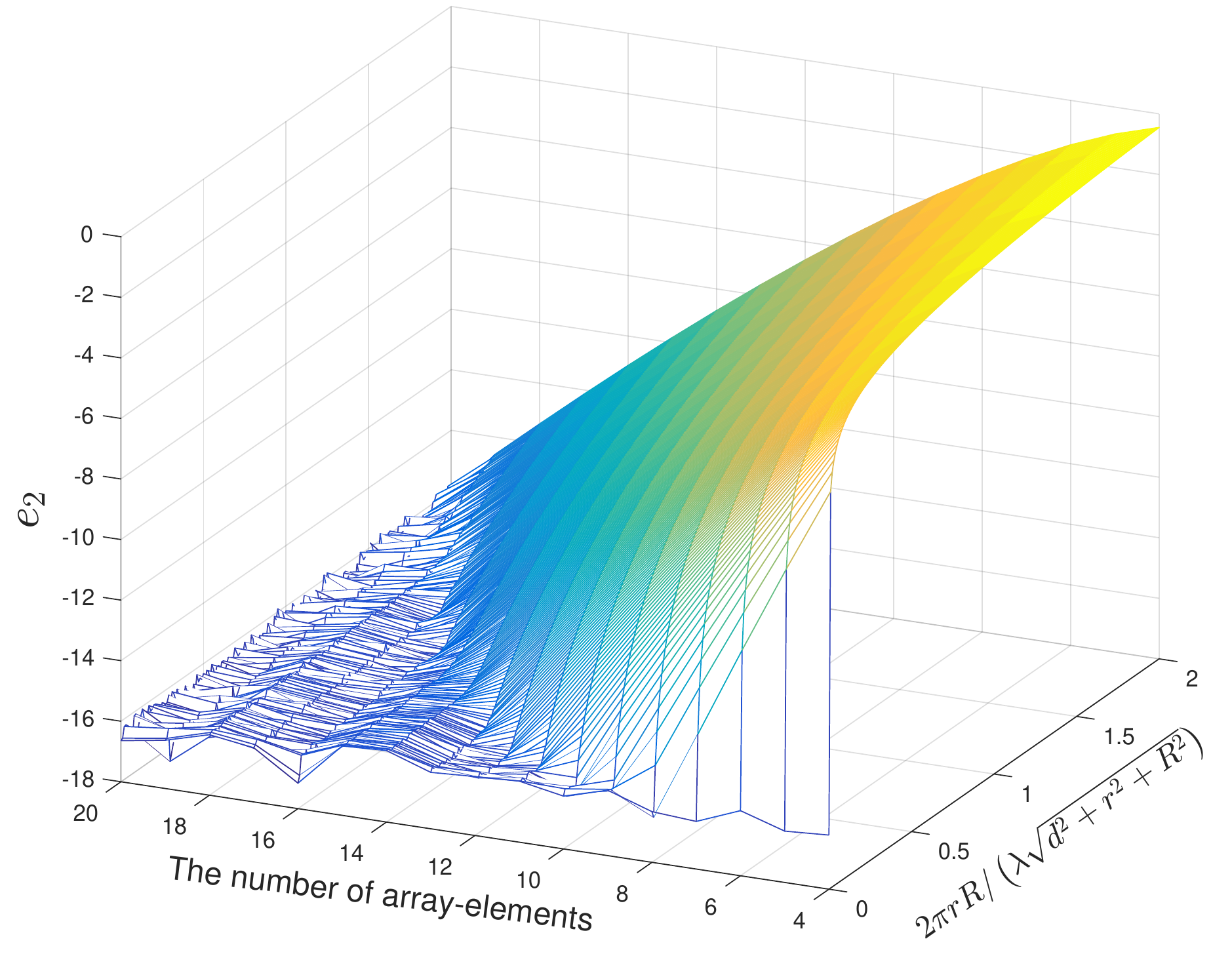}}
\centerline{$l = 2$}
\end{minipage}
\centering
\begin{minipage}[t]{0.32\linewidth}
\centering
\centerline{\includegraphics[width=1\linewidth]{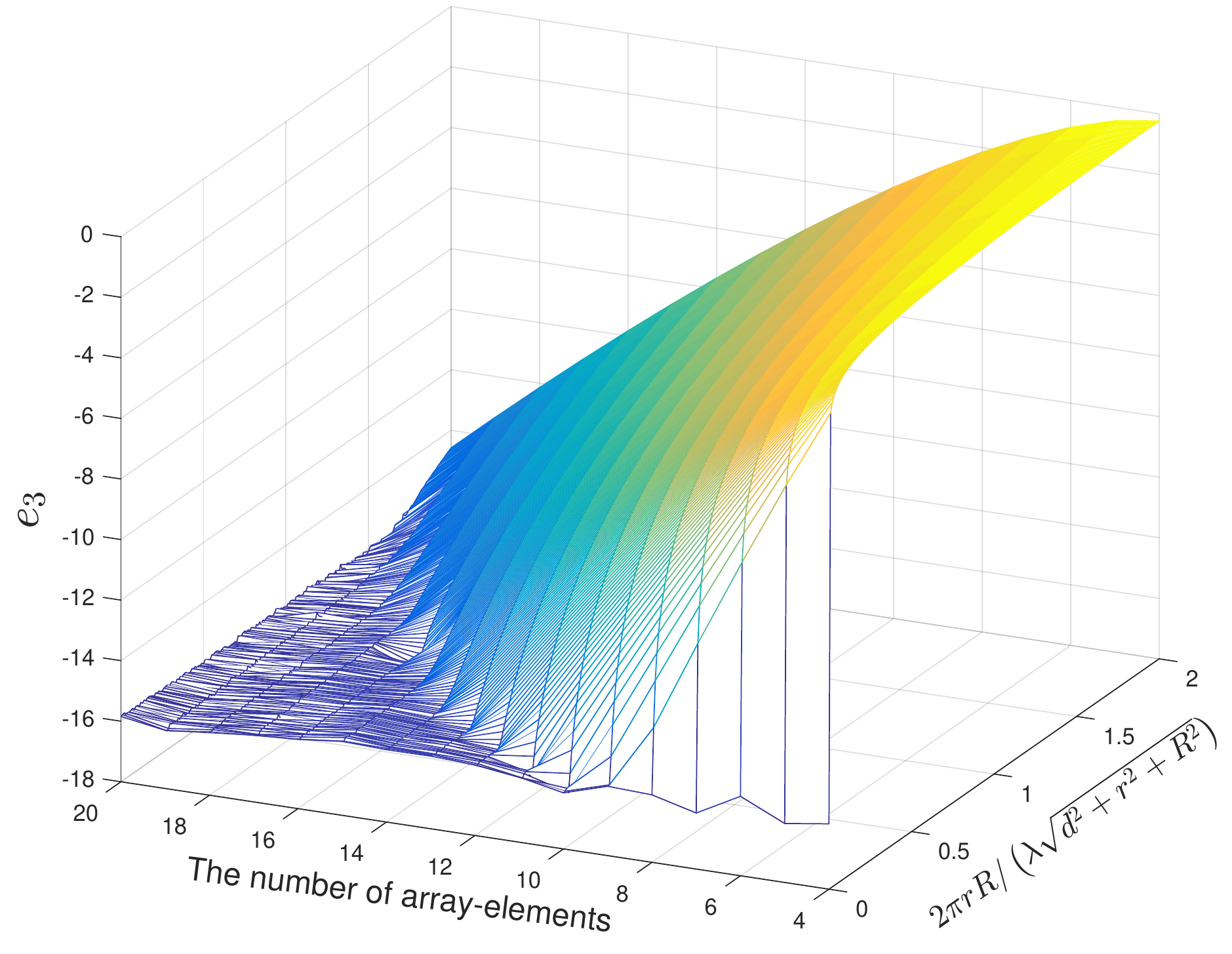}}
\centerline{$l = 3$}
\end{minipage}
\vspace{0.3cm}
\begin{minipage}[t]{0.32\linewidth}
\centering
\centerline{\includegraphics[width=1\linewidth]{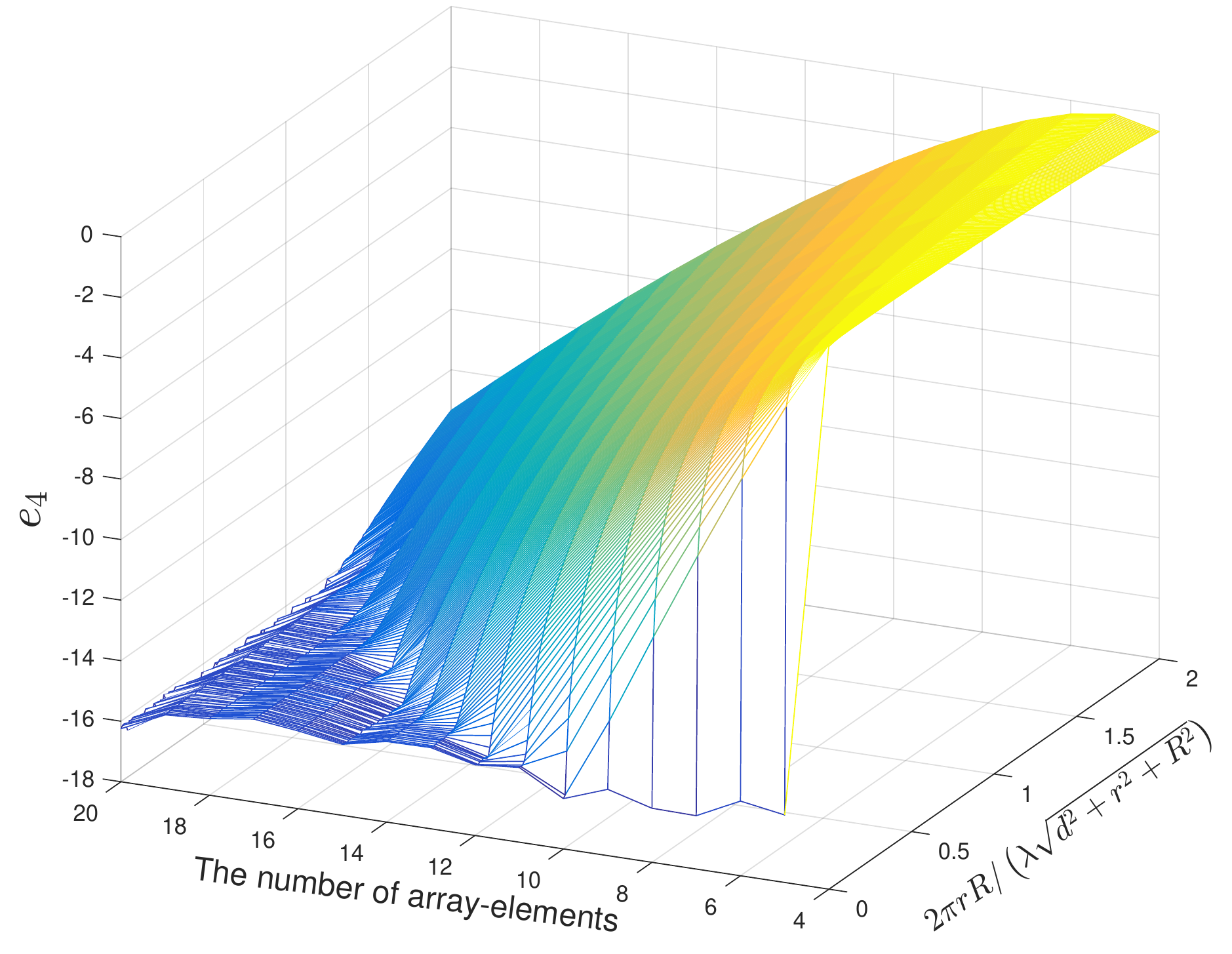}}
\centerline{$l = 4$}
\end{minipage}
\begin{minipage}[t]{0.32\linewidth}
\centering
\centerline{\includegraphics[width=1\linewidth]{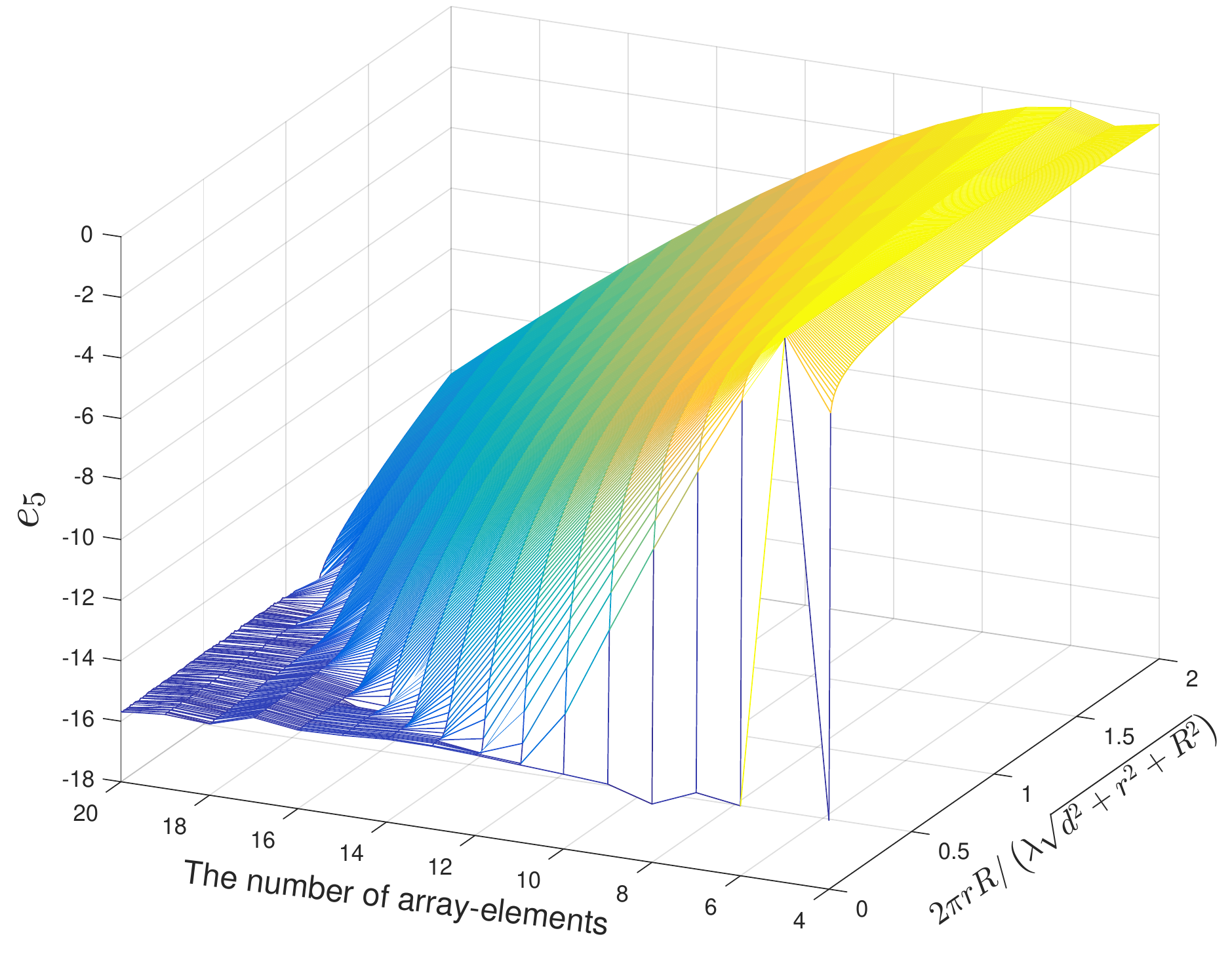}}
\centerline{$l = 5$}
\end{minipage}
\begin{minipage}[t]{0.32\linewidth}
\centering
\centerline{\includegraphics[width=1\linewidth]{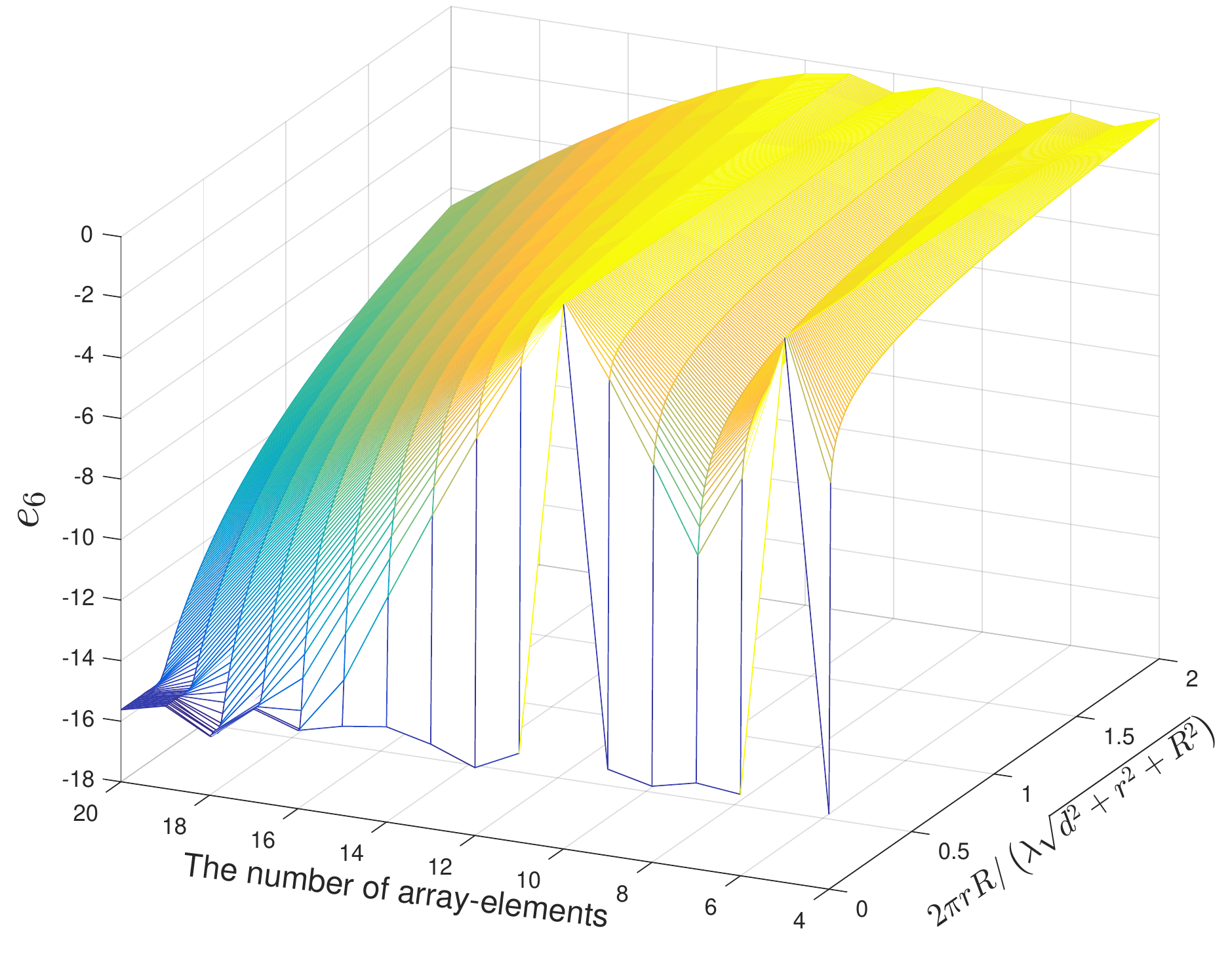}}
\centerline{$l = 6$}
\end{minipage}
\begin{minipage}[t]{0.32\linewidth}
\centering
\centerline{\includegraphics[width=1\linewidth]{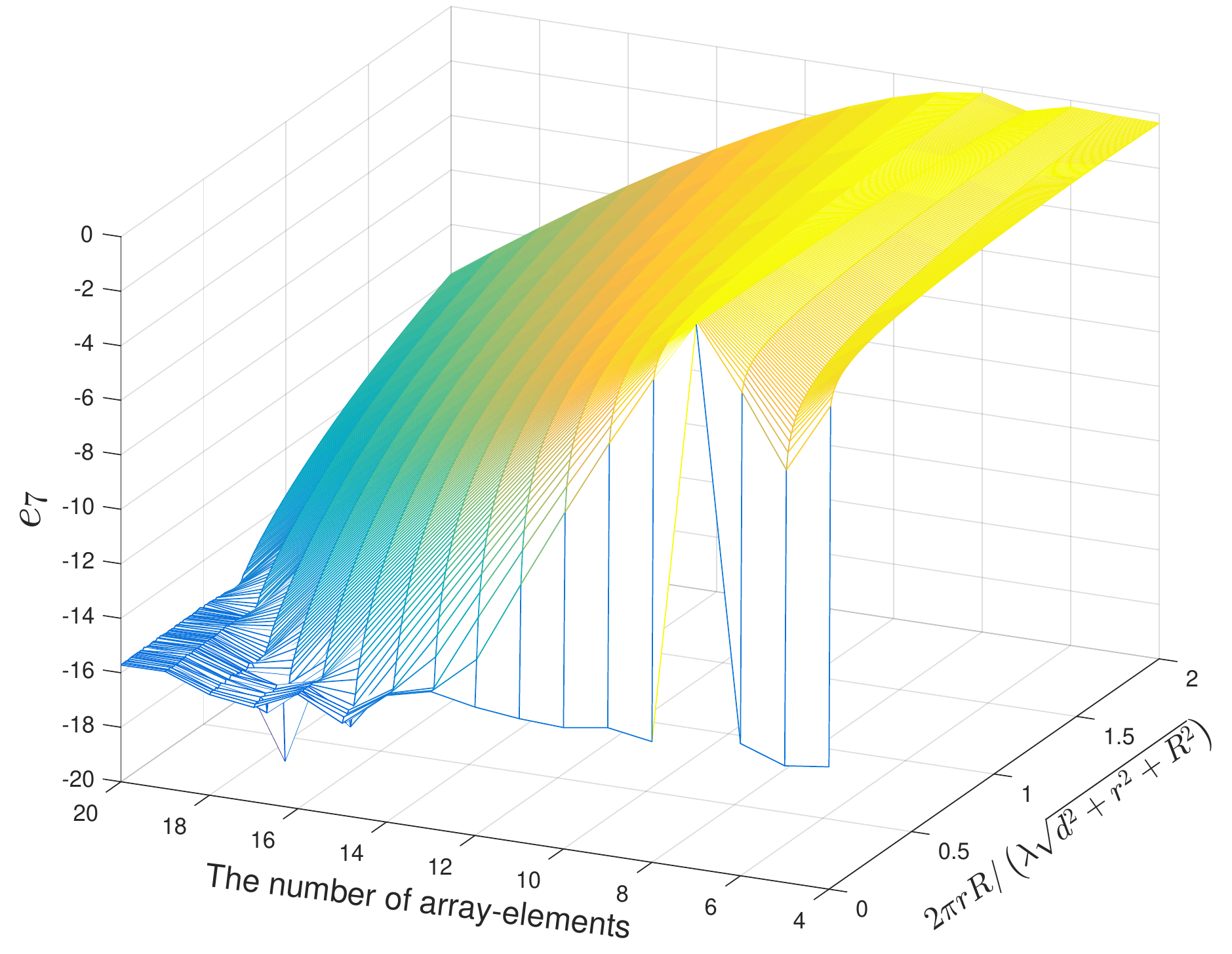}}
\centerline{$l = 7$}
\end{minipage}
\begin{minipage}[t]{0.32\linewidth}
\centering
\centerline{\includegraphics[width=1\linewidth]{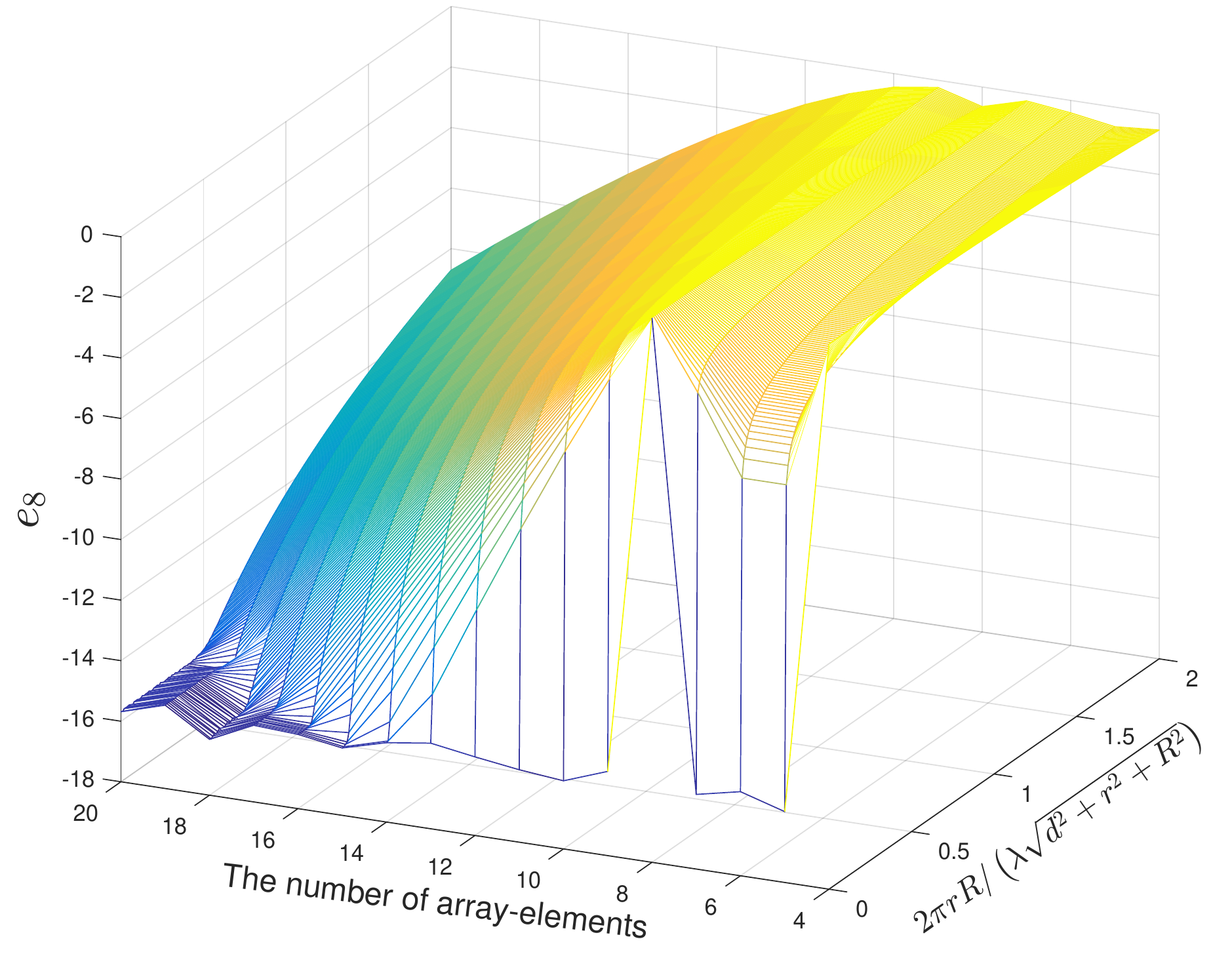}}
\centerline{$l = 8$}
\end{minipage}
    \vspace{8pt}
    \caption{Approximation error from OAM-mode 0 to OAM-mode 8.}
    \label{fig:error}
\end{figure}

Figure~\ref{fig:error} shows the approximation error of different OAM-modes for the non-coaxial UCA transceiver scenario, where we set $\beta=4\pi$, $\alpha_{r}=\alpha_{R}=0$, $\theta=\phi=0$, and $\lambda=0.1$ m. We can observe that when the number of array-elements is larger than 10, the corresponding error is very small and can be ignored. Therefore, $\widetilde{h}_{ml}$ is relatively very accurate when the number of array-elements is larger than 10. Also, the error mainly depends on the number of array-elements while the value of $2\pi r R/\left(\lambda\sqrt{d^{2}+r^{2}+R^{2}}\right)$ has small impact on the error for different OAM-modes.

\begin{figure}
    \centering
    \vspace{0pt}

\centering
\begin{minipage}[t]{0.49\linewidth}
\centering
\centerline{\includegraphics[width=1\linewidth]{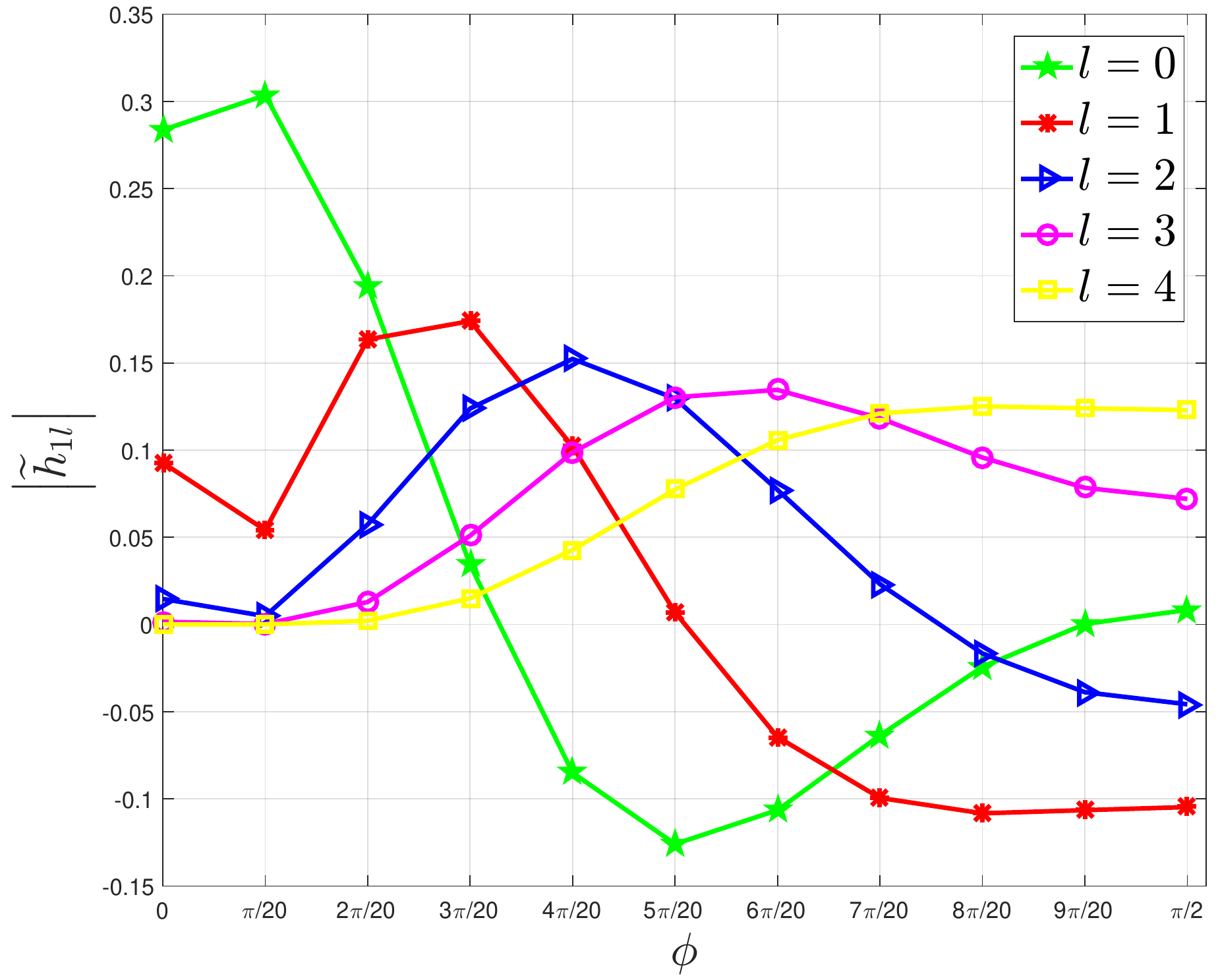}}
\centerline{$m=1$}
\end{minipage}
\centering
\vspace{0.3cm}
\begin{minipage}[t]{0.49\linewidth}
\centering
\centerline{\includegraphics[width=1\linewidth]{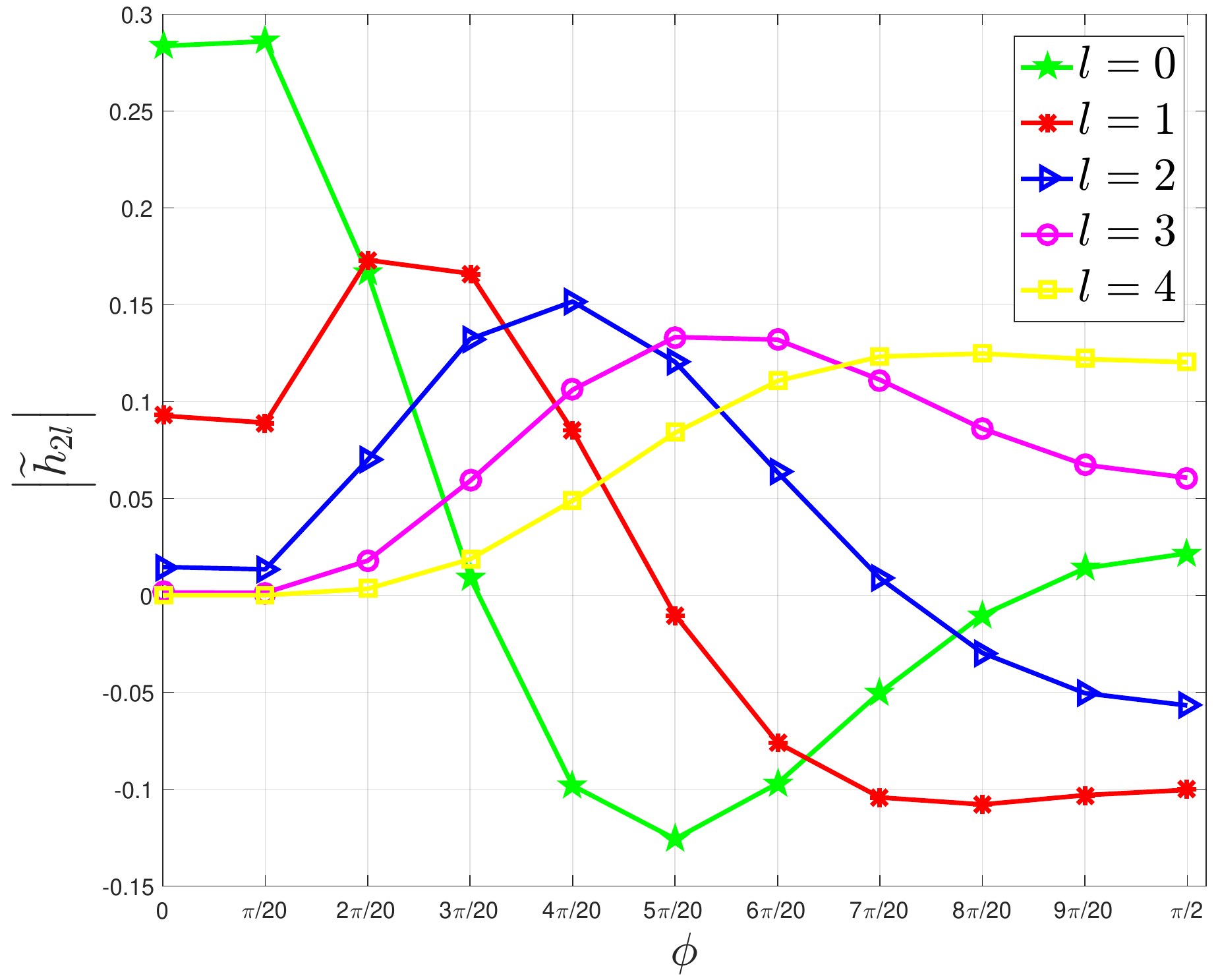}}
\centerline{$m=2$}
\end{minipage}
\centering
\begin{minipage}[t]{0.49\linewidth}
\centering
\centerline{\includegraphics[width=1\linewidth]{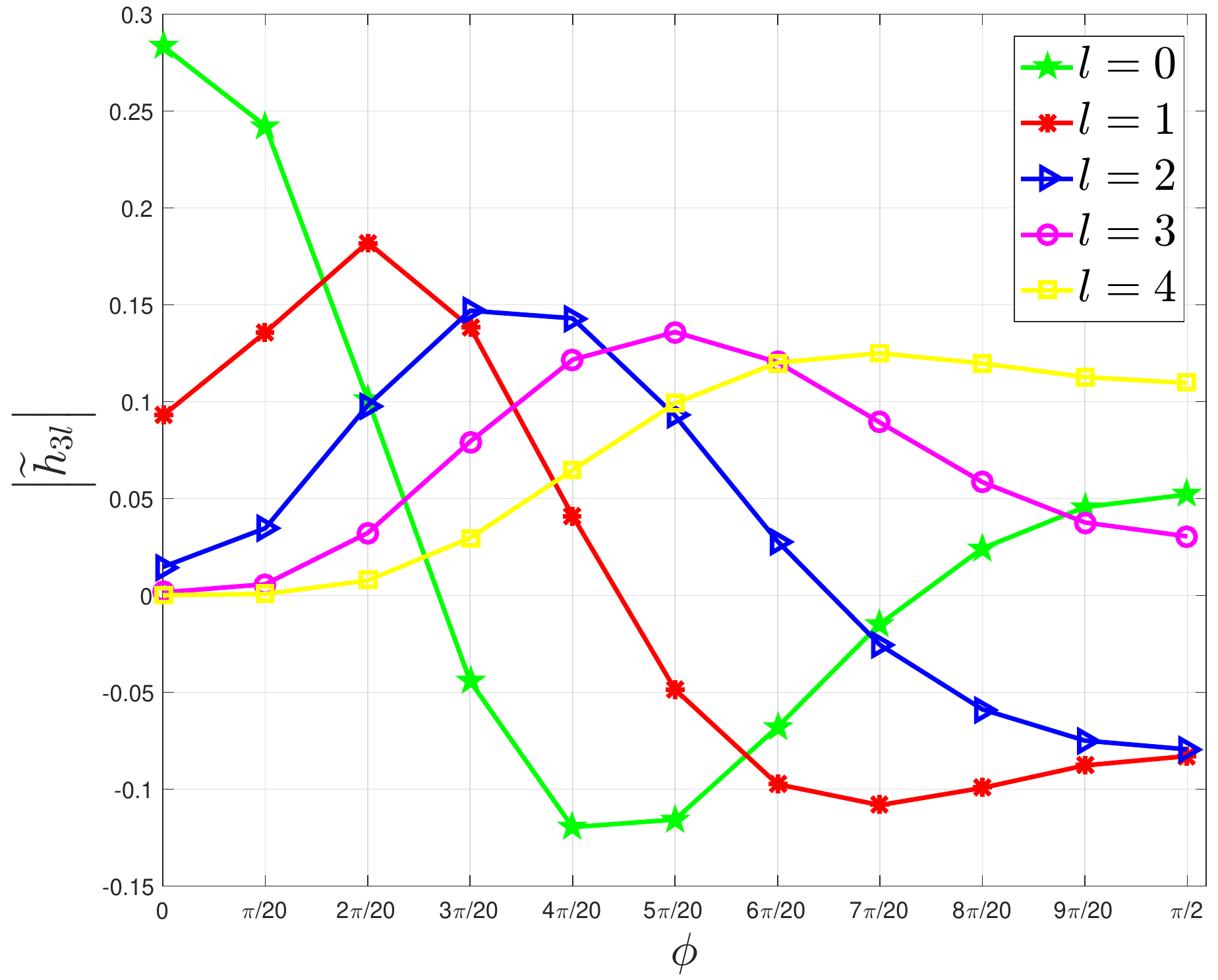}}
\centerline{$m=3$}
\end{minipage}
\centering
\begin{minipage}[t]{0.49\linewidth}
\centering
\centerline{\includegraphics[width=1\linewidth]{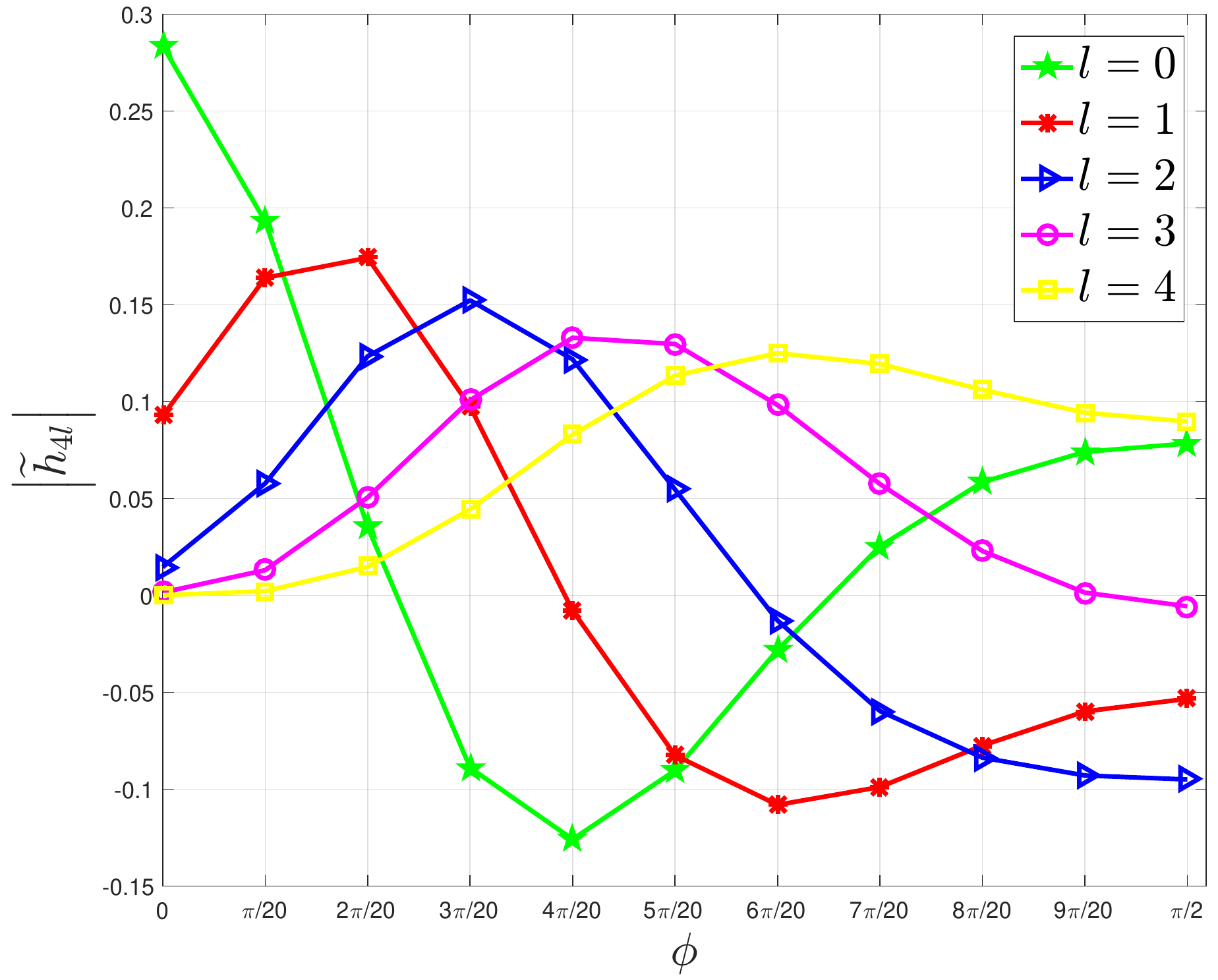}}
\centerline{$m=4$}
\end{minipage}

    \caption{$|\widetilde{h}_{ml}|$ for different OAM-modes with different $m$.}
    \label{fig:h_phi}
\end{figure}

Figure~\ref{fig:h_phi} displays the values of $|\widetilde{h}_{ml}|$ for different OAM-modes with respect to $\phi$, where we set $\beta=4\pi$, $N=M=10$, $\lambda=0.1$ m, $r=R=\lambda$, $d=10\lambda$, $\alpha_{r}=\alpha_{R}=0$, and $\theta=0$. As shown in Fig.~\ref{fig:h_phi}, the channel amplitude gain corresponding to OAM-mode 0 first decreases and then increases as the included angle $\phi$ increases. While the channel amplitude gains corresponding to other OAM-modes first increase and then decrease as the included angle $\phi$ increases. The channel amplitude gains corresponding to different basic angle $\psi_{m}$ are almost the same.

\begin{figure}
    \centering
    \vspace{0pt}

\centering
\begin{minipage}[t]{0.49\linewidth}
\centering
\centerline{\includegraphics[width=1\linewidth]{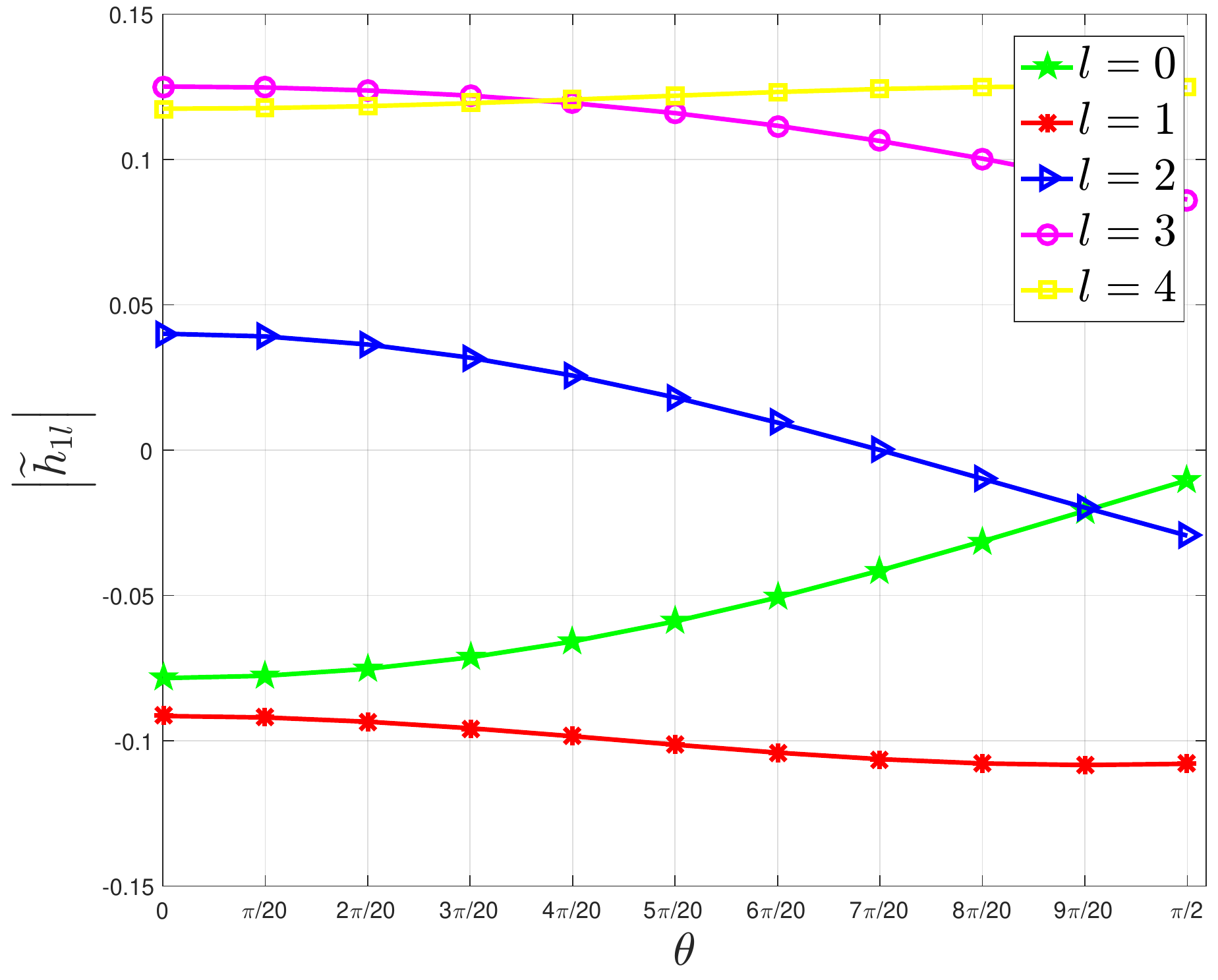}}
\centerline{$m= 1$}
\end{minipage}
\centering
\vspace{0.3cm}
\begin{minipage}[t]{0.49\linewidth}
\centering
\centerline{\includegraphics[width=1\linewidth]{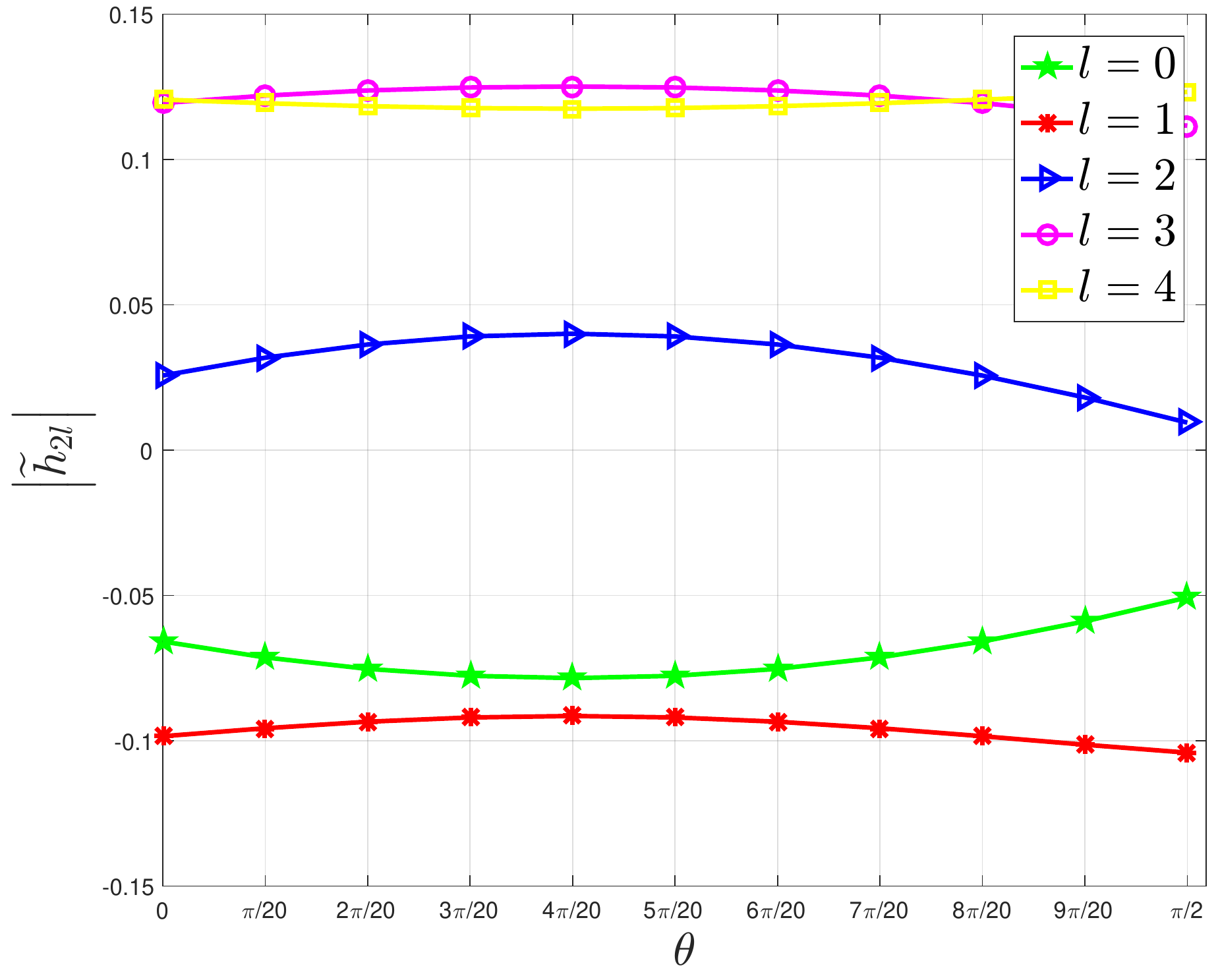}}
\centerline{$m = 2$}
\end{minipage}
\centering
\begin{minipage}[t]{0.49\linewidth}
\centering
\centerline{\includegraphics[width=1\linewidth]{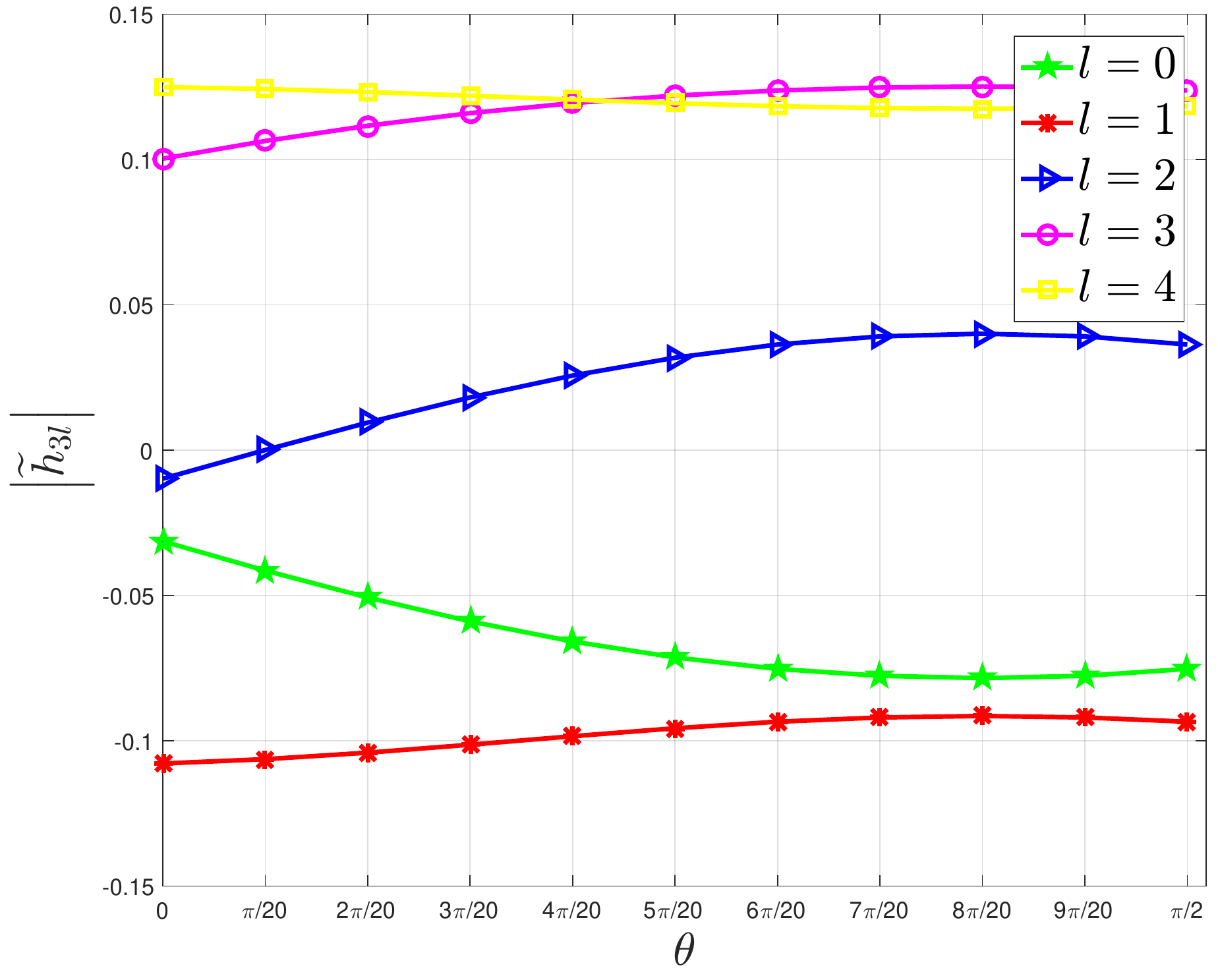}}
\centerline{$m = 3$}
\end{minipage}
\centering
\begin{minipage}[t]{0.49\linewidth}
\centering
\centerline{\includegraphics[width=1\linewidth]{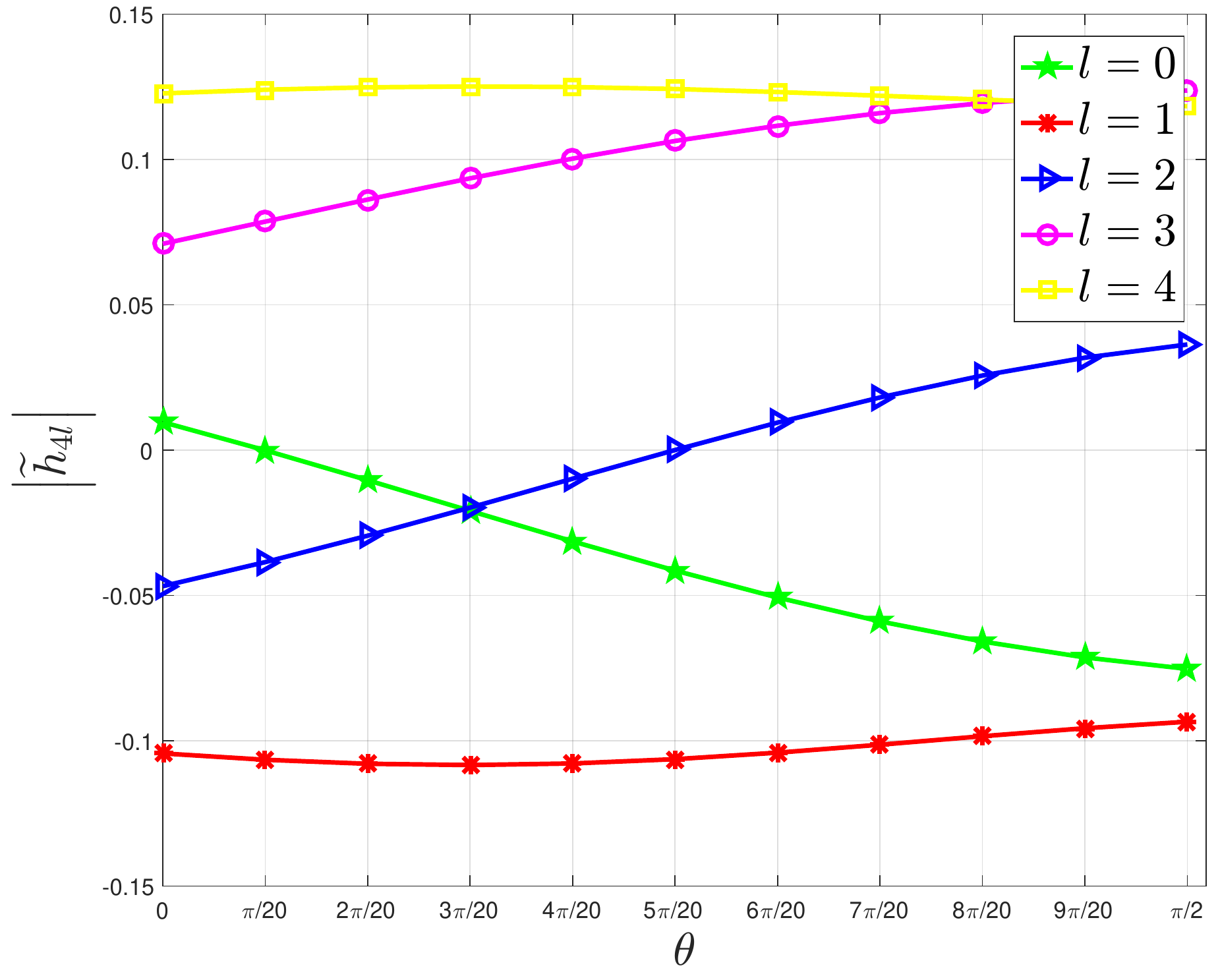}}
\centerline{$m = 4$}
\end{minipage}

    \caption{$|\widetilde{h}_{ml}|$ for different OAM-modes with different $m$.}
    \label{fig:h_theta}
\end{figure}

Figure~\ref{fig:h_theta} depicts the values of $|\widetilde{h}_{ml}|$ for different OAM-modes with respect to $\theta$, where we set $\beta=4\pi$, $N=M=10$, $\lambda=0.1$ m, $r=R=\lambda$, $d=10\lambda$, $\alpha_{r}=\alpha_{R}=0$, and $\phi=\pi/3$. For different $\psi_{m}$, the channel amplitude gains randomly change. Also, the dynamic range of channel amplitude gains for all the OAM-modes is small as the included angle $\theta$ increases.

\begin{figure}
\centering
\includegraphics[width=0.49\textwidth]{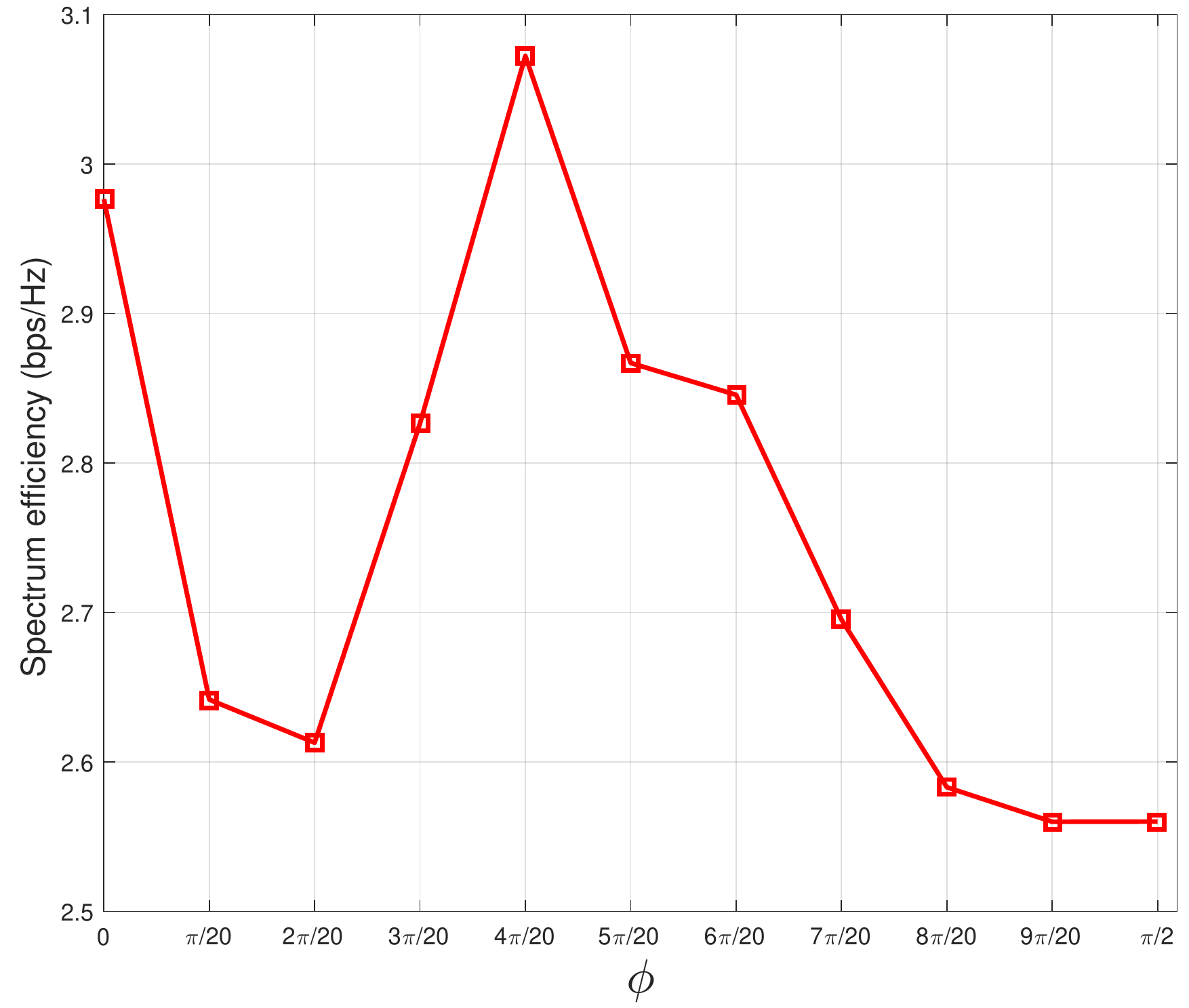}
\caption{The spectrum efficiency versus the included angle $\phi$.}
\label{fig:SE}
\end{figure}
Figure~\ref{fig:SE} shows the spectrum efficiency versus the included angle $\phi$, where we set $\beta=4\pi$, $N=M=10$, $\lambda=0.1$ m, $r=R=\lambda$, $d=10\lambda$, $\alpha_{r}=\alpha_{R}=0$, and $\theta=0$. We can observe that the maximum spectrum efficiency is achieved when the included angle $\phi$ is approximately equal to $2\pi/5$. In some cases, the spectrum efficiency of the non-coaxial UCA transceiver is larger than that of the aligned UCA transceiver based radio vortex wireless communications. That is to say, the scenario where the transmit and receive UCAs are aligned with each other is not the best scene to achieve the spectrum efficiency maximization.

\section{Conclusions} \label{sec:conc}

We investigated the non-coaxial scenario where the transmit and receive UCAs are non-coaxial with each other. Then, we derived the channel model and developed the mode-decomposition scheme to obtain the transmit signal corresponding to each OAM-mode. Then, we discussed the impact of included angles $\theta$ and $\phi$ on the channel gains. The simulation results validated that in some cases the spectrum efficiency of the non-coaxial UCA transceiver is larger than that of the aligned UCA transceiver based radio vortex wireless communications.

\end{document}